\documentclass[a4paper, fleqn, usenatbib]{mnras}

\usepackage{newtxtext,newtxmath}
\usepackage[T1]{fontenc}
\usepackage{ae,aecompl}
\usepackage{soul}

\usepackage{graphicx}	
\graphicspath{{./fig_png/}} 
\usepackage{amsmath}	
\usepackage{color}		
\usepackage{hyperref}	
\usepackage{float}
\usepackage{rotating}
\usepackage{siunitx}

\newcommand{\mpc}{\rm \,cMpc}
\newcommand{\igm}{\rm IGM,\,cell}
\newcommand{\igmcell}{\rm N,\,IGM}
\newcommand{\kpc}{\rm \,kpc}
\newcommand{\Msun}{\rm \,M_{\odot}}
\definecolor{ao(english)}{rgb}{0.0, 0.5, 0.0}
\definecolor{cadmiumgreen}{rgb}{0.0, 0.42, 0.24}
\newcommand{\sentencetoverify}{\color{ao(english)}}

\defcitealias{Mao2019}{Paper I}

\title[Subgrid clumping II]{The impact of inhomogeneous subgrid clumping on
  cosmic reionization II: modelling stochasticity}

\author[M. Bianco et al.]{
	Michele Bianco,$^1$\thanks{Contact e-mail: M.Bianco@sussex.ac.uk} 
	Ilian T. Iliev$^1$,
	Kyungjin Ahn$^2$, 
	Sambit K. Giri$^{3,4}$,\newauthor
	Yi Mao$^5$,
	Hyunbae Park$^6$, 	
	and Paul R. Shapiro$^7$\\
	$^1$Astronomy Centre, Department of Physics \& Astronomy, Pevensey III Building, University of Sussex, Falmer, Brighton, BN1 9QH, UK\\
	$^2$ Department of Earth Science, Chosun University, Gwangju 501-759, Korea\\
	$^3$Institute for Computational Science, University of Zurich, Winterthurerstrasse 190, 8057 Zurich, Switzerland\\
	$^4$Department of Astronomy and Oskar Klein Centre, AlbaNova, Stockholm University, SE-106 91 Stockholm, Sweden\\
	$^5$Department of Astronomy Tsinghua University, Beijing 100084, China\\
	$^6$ Kavli Institute for the Physics and Mathematics of the Universe (WPI), The University of Tokyo Institutes for Advanced Study, The University of Tokyo, \\ Kashiwa, Chiba 277-8583, Japan\\
	$^7$ Department of Astronomy and Texas Cosmology Center, University of Texas, Austin, TX 78	712, USA\\
}

\date{Accepted 2021 March 14. Received 2021 February 8; in original form 2021 February 8}
\pubyear{2021}

\begin{document}
\label{firstpage}
\pagerange{\pageref{firstpage}--\pageref{lastpage}}
\maketitle
\begin{abstract}
Small-scale density fluctuations can significantly affect reionization, but are typically modelled quite crudely. Unresolved fluctuations in numerical simulations and analytical calculations are included using a gas \textit{clumping factor}, typically assumed to be independent of the local environment. In Paper I we presented an improved, local density-dependent model for the sub-grid gas clumping. Here we extend this using an \textit{empirical stochastic model} based on the results from high-resolution numerical simulations  which fully resolve all relevant fluctuations. Our model reproduces well both the mean density-clumping relation and its scatter. We applied our stochastic model, along with the mean clumping one and the Paper I deterministic model, to create a large-volume realisations of the clumping field, and used these in radiative transfer simulations of cosmic reionization. Our results show that the simplistic mean clumping model delays reionization compared to local density-dependent models, despite producing fewer recombinations overall. This is due to the very different spatial distribution of clumping, resulting in much higher photoionization rates in the latter cases. The mean clumping model produces smaller H~II regions throughout most of reionization, but those percolate faster at late times. It also causes significant delay in the 21-cm fluctuations peak and yields lower non-Gaussianity and many fewer bright pixels in the PDF distribution. The stochastic density-dependent model shows relatively minor differences from the deterministic one, mostly concentrated around overlap, where it significantly suppresses the 21-cm fluctuations, and at the bright tail of the 21-cm PDFs, where it produces noticeably more bright pixels.

\end{abstract}

\begin{keywords}
Cosmology: theory, dark ages, reionization, first stars -- Methods: numerical -- Galaxies: intergalactic medium
\end{keywords}

\section{Introduction}
The Epoch of Reionization (EoR) is an important period in the history of the Universe, which encompasses the creation of the first stars and galaxies that subsequently influenced the formation and evolution of latter-day structures. These luminous objects have produced enough UV-radiation to both alter their host galaxy composition and to propagate into the intergalactic medium (IGM), ultimately ionizing it for a second time \citep{Furlanetto2006, Zaroubi2012, Ferrara2014}.

The key goal of reionization simulations is to provide numerical framework for constraining EoR observables, for example the detection of the 21cm hyperfine transition of neutral hydrogen fluctuations \citep{Bowman2010, Paciga2013, Yatawatta2013, Parsons2014, Jelic2014, Jacobs2015, Dillon2015, Robertson2015, Ali2015, Pober2015, Patil2017, Mertens2020, Ghara2020} and Lyman-$\alpha$ damping wings \citep{Davies2018, Greig2019}. Such simulations require large volumes, of several hundreds $\rm cMpc$ size in order to correctly derive the cosmic reionization history, to account for abundance and clustering of expected sources and to sample vast regions of the universe for detection of the redshifted 21cm hyperfine transition of neutral hydrogen \citep{Mellema2006, Iliev2014}, as relevant for current and upcoming experiments (e.g. LOFAR\footnote{\url{http://lofar.org}}, SKA\footnote{\url{https://skatelescope.org}}). At the same time, EoR simulations need to include fluctuations in the density distribution down to the Jeans mass of the cold gas, which is in sub-$\rm kpc$ scale, so as to correctly model recombination effects and thus properly track the expansion of ionizing fronts throughout reionization \citep{Shapiro2018}. Unfortunately, because of limited dynamic range, satisfying both of these requirements in a single fully-numerical simulation is currently unachievable, and will remain challenging in the future. Hence, in large-scale simulations, the sources and sinks of ionizing radiation often act on scales much smaller than the resolution level and need to be treated using sub-grid prescriptions. Consequently, simulations may adopt incorrect values for various relevant quantities (e.g. density, temperature, gas pressure, etc.) smoothed on the (relatively coarse) grid scales and this could influence the predicted observational signatures.

In this work, our focus is on how sub-grid density inhomogeneities are considered within the volume elements of large-scale simulations. Depending on how gas density fluctions vary in space and over time (local degree of \textit{"clumpiness"}), the recombinations in the IGM can significantly affect the progress and nature of the reionization process. For every ionized atom that recombines with a free electron, an additional ionizing photon should be produced in order to ionize it again and keep the IGM highly ionized. In this way potentially a substantial portion of the sources photon budget could be depleted. In simulations, the recombination rate $\mathcal{R}$ is discrete, averaged on a mesh giving $\rm \left< \mathcal{R} \right> = \left<\alpha_B(T) x^2_i\, n^2 \right>$, where $\alpha_B(T)$ is the (temperature-dependent) Case B recombination coefficient, $x_i$ is the ionized fraction, $n$ is the number density and for simplicity we assumed pure hydrogen gas. This indicates the number of electron-proton recombination per second in a volume, for a given gas chemistry, within each grid cell. Early semi-analytical models have adopted a common methodology named \textit{Clumping Factor Approach}, that defines the averaged recombination rate in terms of a \textit{clumping factor} $\rm C = \left< n^2\right> / \left< n \right>^2 $, which corrects for the difference between the cell-averaged $\langle n\rangle^2$ and the actual value, thereby accounting for unresolved small-scale (sub-grid) structure in simulations \citep{Gnedin1997, Tegmark1996, Ciardi1997, Madau1999, Valageas1999}. If not correctly treated, this approach can underestimate the impact of sub-grid inhomogeneities on absorption of radiation. In some cases this term is just completely ignored, i.e.: $\rm C=1$ \citep{Onken2004, Kohler2007}, but the more common and simplistic approaches consist in either a constant global term \citep{Cen2003, Zhang2007} or a time evolving global term \citep{Iliev2005a, Mellema2006, Iliev2007, Pawlik2009}, averaged on the entire box volume, also referred as the \textit{biased homogeneous} or \textit{globally averaged clumping model}. Recently we presented our first work \cite[see][for reference]{Mao2019}, hereafter Paper I, where we investigated the impact of a spatially varying, local density dependent sub-grid clumping factor on reionization observables. In the present paper we extend the discussion and propose a more realistic and accurate treatment of the \textit{Clumping Factor Approach}, that takes into account also the scatter around the mean clumping-density relation observed in high-resolution simulations.

We use a high-resolution N-body simulation of a small volume of side length $9\mpc$, with spatial and mass resolution of approximately $\rm200\,pc$ and $\rm5000\,M_\odot$, to statistically describe IGM density fluctuations down to the Jeans mass in the cold, pre-reonization gas and then to implement these sub-grid density fluctuations into two large volume ($714\mpc$ and $349\mpc$ os side length) reionization simulations. By adapting the small-scale sub-grid to the resolution of larger boxes we then model the correlation between density and clumping factor, comparing three different models (details in \S\ref{sec:ClumpMod}), in order to infer the clumping factor from the coarse density grid of the large volume, see \S\ref{sec:ClumpRealiz}. Finally we perform a radiative transfer simulation to study the effect of this sub-grid inhomogeneity approach on observables of reionization.

This paper is organized as follows. In \S~\ref{chap:Method} we present the N-body and radiative transfer (RT) simulation used, the numerical methods, \S\ref{sec:CoarseGridMeth} and our models in \S\ref{sec:ClumpMod}. In \S\ref{sec:ClumpRealiz} we discuss the realisation of the clumping factor for large volumes from sub-grid inhomogeneity correlation. In \S\ref{chap:RTresult} we analyse our RT simulation results and look into how our models influence the basic features of EoR: the reionization history in $\S$\ref{sec:IonizHistory}, the volume-averaged ionization fraction evolution, the integrated Thompson optical depth and then the Bubble size distribution in $\S$\ref{sec:BubbleSize}. To better understand the change in ionization morphology we describe a side-by-side comparison of box slice shot with zoom $\S$\ref{sec:SliceComparison}. In \S\ref{sec:21cmSign} we analyse the 21cm signal power spectra and the brightness temperature distribution. Our conclusions are summarized in \S\ref{chap:Conclusion}.

\section{Methodology} \label{chap:Method}
\subsection{Numerical Simulations} \label{sec:Nbody}

\begin{table*}
	\begin{center}
		\begin{minipage}{1\linewidth}
			\caption{N-body simulation parameters. Minimum halo mass is $10^5\,M_\odot$, $10^9\,M_\odot$ and $10^9\,M_\odot$, corresponding to 20, 40 and 25 particles, respectively in SB, LB-1 and LB-2. In all cases the force smoothing length is fixed at $1/20$ of the mean inter-particle spacing.}
			\label{tab:simulation}
				\begin{tabular}{cccccccc} \hline
				Label	&	Box size	&	$N_{particle}$	&	fine mesh	&	spatial resolution	&	$m_{particle}$			& RT coarse-grained mesh\footnote{SB density grid is coarsened to the  to the required resolution for the LBs. In the column for SB, the coarsened mesh size and respective percentage of the overlapping volume for windows mesh function, calculated with \autoref{eq:noc}.} 	& RT coarse-grained cell size\footnote{Spatial resolution of the RT coarse-grained mesh for SB, for the calculation of \autoref{eq:n_cell} and \ref{eq:n2_cell}.} \\\hline\hline
				SB		&	$9\mpc$		&	$1728^3$		&	$3456^3$	&	$260\,$pc			&	$5.12\times10^{3}\Msun$	& $8^3\,(53\%),13^3\,(50\%)$ & $2.381,1.394\mpc$\\
				LB-1	&	$714\mpc$	&	$6912^3$		&	$13824^3$	&	$5.17\kpc$			&	$4.05\times10^{7}\Msun$	& $300^3$					& $2.381\mpc$\\ 
				LB-2	&	$349\mpc$	&	$4000^3$		&	$8000^3$	&	$4.36\kpc$			&	$2.43\times10^{7}\Msun$	& $250^3$					& $1.394\mpc$\\\hline
			\end{tabular}
		\end{minipage}
	\end{center}
\end{table*}

We use N-body simulations to follow the evolution of cosmic structures, performed with the \texttt{CUBEP$^3$M} code \citep{Harnois-Deraps2013}. The code uses particle-particle on short-range and particle-mesh on long-range to calculate gravitational forces. We use set of three N-body simulations, whose parameters are summarized in \autoref{tab:simulation}. 

Our clumping factor modeling is based on small, high resolution volume box ($6.3\,h^{-1}\,$Mpc$=9\,$Mpc, $1728^3$ particles, labelled SB in \autoref{tab:simulation}). This has sufficient spatial and mass resolution to resolve the smallest halos that can hold cold, neutral gas. Our main larger-volume N-body simulation is referred to as LB-1 ($500\,h^{-1}$Mpc$=714\,$Mpc, $6912^3\approx330$~billion particles). A smaller simulation, LB-2, ($244\,h^{-1}$Mpc$=349\,$Mpc, $4000^3=64$~billion particles) will be used as comparison to analyse possible influence of box size and resolution in the realisation of sub-grid clumping factor and prove the stability of our method. For both of the large-volume simulations the minimum halos mass resolved is $10^9\,M_\odot$, while halos with $10^8\,M_\odot<M_{halo}<10^9\,M_\odot$ are implemented using a sub-grid model \citep{Ahn2015}, thereby all atomically-cooling halos (ACHs) with minimum mass $\rm M_{halo} \gtrsim 5 \times 10^8 \Msun$ are included. We are using updated N-body simulations compared to \citetalias{Mao2019}, we illustrate this further in \S\ref{sec:oldnew}.

An on-the-fly spherical overdensity halo finder \citep{Harnois-Deraps2013,Watson2013}, with overdensity parameter $\Delta=130$, creates an halo catalogues at given redshift, that is later used as inputs for the radiative transfer simulation. The remaining particles are categorized as part of the IGM. In this work we do not include any effects from minihaloes $\rm M_{halo}<10^{8}\Msun$. Even though these sources could have driven ionization in the early phase of EoR, their effect on later stage is expected to be minor because of molecular dissociation by UV background radiation from primordial luminous sources, up to a point that their contribution is negligible compared to heavier ACHs \citep{Ahn2009}. Initial conditions are generated using the Zel'dovich approximation and the power spectrum of the linear fluctuations is given by the \texttt{CAMB} code \citep{Lewis2000}.
The SB N-body simulation starts at redshift $\rm z=300$, while LB-1 and LB-2 at $z=150$, which gives enough time to significantly reduced non-linear decaying modes \citep{Crocce2006}, while at the same time fluctuations are small enough to ensure linearity of density field at the respective resolutions. The cosmological parameter are based on WMAP 5 years data observation and consistent
with final Planck results, for a flat, $\Lambda$CDM cosmology with the following parameters, $\rm\Omega_{\Lambda}=0.73$, $\rm\Omega_{m}=0.27$, $\rm\Omega_b=0.044$, $\rm H_0=70\,km\,s^{-1}\mpc^{-1}$, $\rm\sigma_8=0.8$, $\rm n_s=0.96$ and the cosmic helium abundance $\rm\eta_{He}=0.074$ \citep{Komatsu2010}. Our method is general and can be applied in any cosmological background, but the specific fitting parameters we provide are based on the above values.\\

We simulate the Epoch of Reionization using the \texttt{C$^2$-Ray} code \citep{Mellema2006}, a photon-conserving radiative transfer (RT) code based on short characteristic ray-tracing. The LB-1 and LB-2 N-body simulations provide the IGM density fields and halo catalogues with masses, velocities, position and other variables, for a total of $76$ snapshots, equally spaced in time ($\Delta t=11.54\,$Myr) in the redshfit interval $\rm z\in[6;\,50]$. For computational feasibility, the density grid is coarsened for the radiative transfer simulation to $300^3$ (LB-1), and $250^3$ (LB-2). The high-resolution N-body simulation (SB) data input is initially interpolated onto a $1200^3$ (SB) grid, which can then be coarsened to the required resolution as discussed in the next section. These grids correspond respectively to cell sizes of length $2.381 \mpc$, $1.394 \mpc$ and $7.5 \kpc$. For brevity we will refer to these grids as the \textit{sub-grid volumes} for SB, and \textit{coarse volumes} in LB-1 and LB-2, noted $\left< . \right>_{crs}$. Just as in \citetalias{Mao2019}, the interpolation of the particles onto a grid is performed with a Smoothed-Particle-Hydrodynamic-like method (SPH-like), which then yields coarse-grid density, velocity and clumping fields (see sect. 2.2 in Paper I
for details).

Ionization sources for the radiative transfer simulations are characterised by the ionizing photon production rate per unit time $\dot{N}_{\gamma}$, given by
\begin{equation} \label{eq:dotNgamma}
	\dot{N}_{\gamma} = f_{\gamma} \frac{M_{halo}\,\Omega_b}{\Delta t_s m_p \Omega_0}
\end{equation}
where $m_p$ is the proton mass, $M_{halo}$ is the total halo mass within coarse-grid cell, $\Delta t_s=11.53\,$Myr, the lifetime of stars set equal to the time between N-body snapshots. $\rm f_{\gamma}$ is the efficiency factor, defined as
\begin{equation} \label{eq:fgamma}
	 f_{\gamma} = f_{\star}\, f_{esc}\, N_i
\end{equation}
where $f_{\star}$ is the star formation efficiency, $f\rm_{esc}$ is the photons escape fraction and $N_i$ is the stars ionizing photon production efficiency per stellar atom, it depends on the initial mass function (IMF) of the stellar population, e.g. for Pop II (Salpeter IMF) $\dot{N}_{\gamma} \sim 4000$, the value for $f_{\star}$ and $f_{esc}$ are still uncertain, therefore these parameters can be tuned in order to match the observational constrain that we will discuss in $\S$\ref{chap:RTresult}. Here we adopt the partial suppression model of \citep{Dixon2015}, whereby for LMACHs located in a neutral cell the efficiency factor is set to $f\rm_{\gamma}=8.2$, while in an ionized cell (above 10\%) we set $f\rm_{\gamma}=5$ to account for feedback. For HMACHs the efficiency factor has a constant value of $f\rm_{\gamma}=5$, equivalent to e.g. $\rm N_i=5000$, $f\rm _{\star}=0.05$ and $f\rm_{esc}=0.02$.\\

\subsection{Coarse-Grid Method} \label{sec:CoarseGridMeth}
Our clumping factor calculations are based on N-body data and neglect any hydrodynamical effects on the clumping factor. Accounting for the gas pressure provides additional smoothing at small scales and therefore our clumping factor values should be considered as upper limits. Moreover, we are interested in the reionization of the IGM and therefore exclude the halos from our calculations. The contribution of recombinations inside haloes is already taken into account in \autoref{eq:dotNgamma} through the photon escape fraction and should not be counted again.

In order to represent the N-body particles into a regular grid, we adopt the SPH-like smoothing technique described in $\S${\color{blue}2.2} of \citetalias{Mao2019}, we refer the readers to e.g. \cite{Shapiro1996} for more general details on SPH smoothing methods. In
LB-1 and LB-2 simulations we use regular meshes directly produced by the SPH code at the required resolution (the specific values used here are listed in \autoref{tab:simulation}). In the SB simulation we adopt a more flexible approach, whereby we first produce all quantities on a very fine mesh (here $1200^3$), which is later coarsened as required in order to approximately match the cell sizes used in the LB simulations.

\begin{figure*}
	\centering
	\makebox[\linewidth]{\includegraphics[scale=0.2]{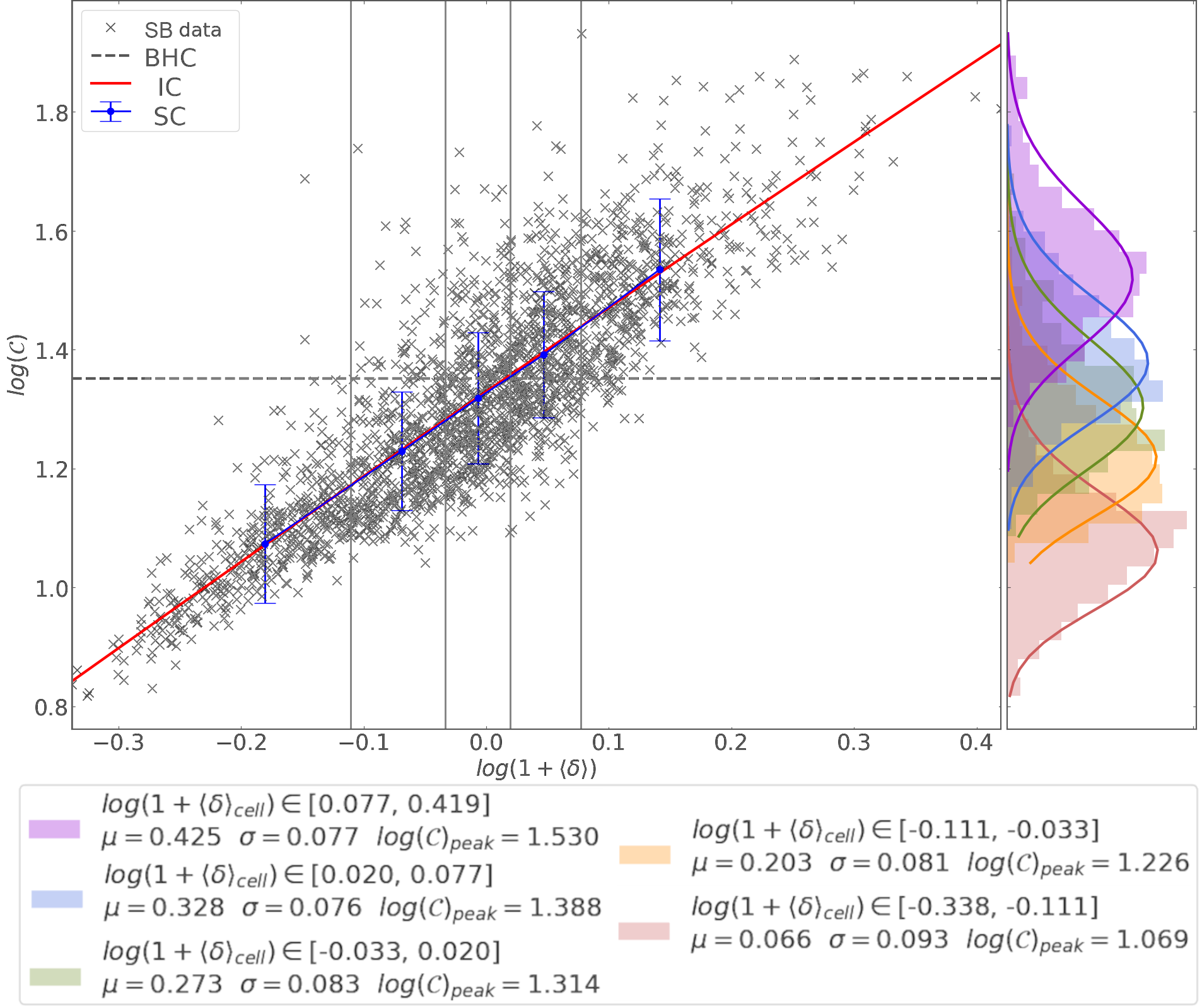} \includegraphics[scale=0.2]{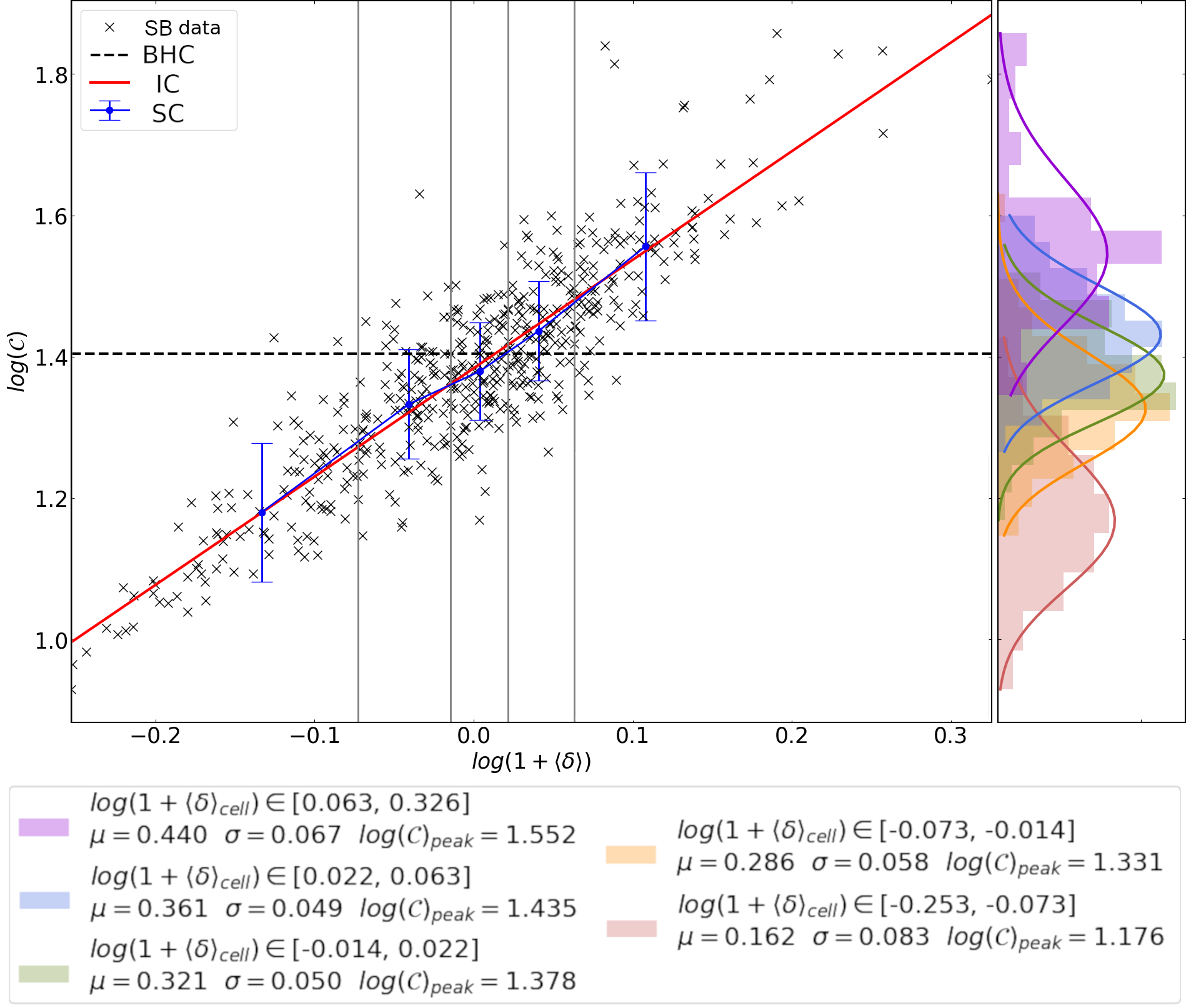}}\vskip-2mm
	\caption{Sample correlation between local coarse IGM overdensity and coarse clumping factor at redshift $\rm z=7.305$ for LB-1 resolution (1.394 Mpc cells, left panels) and LB-2 resolution (2.391 Mpc, right). Shown are the coarsened SB N-body data at these resolutions (black crosses), the IC model (deterministic) fit (red line) and the globally-averaged clumping factor (horizontal dashed line). The (blue) points with error bars represent the expected value and standard deviation of the log-normal distribution (see text) in each overdensity bin. Vertical lines (solid grey) indicate the bin limits, whose sizes are adjusted so that each bin contains the same number of data points (approximately $\sim 400$ (left) and $100$ (right)). For each figure the right panel shows the log-normal distribution (solid line) of the clumping within each density bin vs. the actual data (shadow area), where we include in the legend below each panels a short description of the relevant parameters. 
	}\label{fig:corr+pdf}
\end{figure*}

A window mesh function smooths the SB mesh-grid on a coarser-grained mesh, with size defined by \autoref{eq:noc}. The method allows the windows function to overlap. The percentage of overlap $\rm N_{\%}$ is chosen in order to achieve the required resolution size of the LBs and at the same time obtain a large enough set of coarsened SB data, since $\rm Mesh_{crs-gr}^3$ gives the total number of data point then interpolated by the clumping models (see \autoref{fig:corr+pdf}).
\begin{equation} \label{eq:noc}
	\rm Mesh_{crs-gr} = \frac{BoxSize_{SB}}{(1-N_{\%})\cdot Res_{LB}}
\end{equation}
where $\rm Res_{LB}$ is the coarse grained resolution of the large box and $\rm BoxSize_{SB}$ the box size of the small box.
We employ the SB cell-wise quantities expressed with equations \eqref{eq:clump_cell} and \eqref{eq:overdens} to compute the parametrization of the correlation models. Hereafter, we will refer to them as the \textit{sub-coarse-grid} or \textit{SB data}, whereas in the case of LBs we name them \textit{RT-mesh grid}. 
In our case we have $\rm Mesh_{crs-gr}=8$ with percentage overlap $\rm N_{\%}=53\%$ for $\rm714\mpc$ (LB-1) and $\rm Mesh_{crs-gr}=13$ with  $\rm N_{\%}=50\%$ for $349\mpc$ (LB-2).

We define the gas clumping factor based on the cell-wise averaged quantities \citep[e.g.][]{Iliev2007, Akila2018, Mao2019}
\begin{equation} \label{eq:clump_cell}
	\rm C_{\igm} = \frac{\langle n^2_{\igmcell} \rangle_{cell}}{\langle n_{\igmcell} \rangle_{cell}^2}
\end{equation}
where 
\begin{equation} \label{eq:n_cell}
	\rm \langle n_{\igmcell} \rangle_{cell} \equiv \frac{1}{V_{cell}} \int_{cell} n_{\igmcell}(\textbf{r})\,d^3r
\end{equation}
and 
\begin{equation} \label{eq:n2_cell}
	\rm \langle n^2_{\igmcell} \rangle_{cell} \equiv \frac{1}{V_{cell}} \int_{cell} n^2_{\igmcell}(\textbf{r})\,d^3r.
\end{equation}
The mean cell over-density is defined
\begin{equation} \label{eq:overdens}
	\rm 1+ \langle \delta \rangle_{cell} = \frac{\langle n_{\igmcell} \rangle_{cell}}{\overline{n}_{\igmcell}}
\end{equation}
where $\rm \overline{n}_{IGM}$ is the global average of the IGM number density over the entire box volume (in this paper, we always refer to quantities in comoving units).\\
\subsection{Modeling the Overdensity-Clumping Correlation} \label{sec:ClumpMod}

In this work we consider several models for the parametrisation of the correlation between the local coarse overdensity $\rm 1+ \langle \delta \rangle_{cell}$ and the coarse clumping factor $\rm C_{\igm}$.  

\subsubsection*{I) Biased Homogeneous Subgrid Clumping (BHC)}
The simplest approach is to set a constant (redshift-dependent) clumping factor $C(z)$, for the entire simulation volume  \citep[e.g.][]{Madau1999, Mellema2006, Iliev2007, Kohler2007, Raicevic2011}.
In our case, we evaluate this globally averaged clumping for every SB simulation snapshot at the appropriate coarse resolution and then fit it with an exponential function of the form:
\begin{equation} \label{eq:Cmean}
	\rm C_{BHC}(z) \equiv \overline{C}_{\igm} =  C_0\,e^{c_1\,z + c_2\,z^2}+1
\end{equation}
where $\rm C_0$, $\rm c_1$ and $\rm c_2$ are the fitting parameters. We refer to this
model as biased homogeneous clumping \citepalias{Mao2019} since that volume-averaged value is then multiplied by the local cell density to obtain the recombination rate, effectively biasing
recombinations towards high-density regions.


\subsubsection*{II) Inhomogeneous Subgrid Clumping (IC) Model}
This model, where the local gas clumping is set based on one-to-one, deterministic relation with the cell density, was first presented in \citetalias{Mao2019}. We include it here for comparison purposes. The relation of the clumping with the overdensity in \autoref{eq:clump_cell} is fit by a quadratic function: 
\begin{equation} \label{eq:quad}
  \rm log_{10}(C_{IC}(x\,|\,z_i)) \equiv y = a_i \, x^2 + b_i\, x + c_i
\end{equation}
where $\rm x=\log_{10}(1+\langle\delta\rangle_{cell})$ and $\rm y=\log_{10}(C_{\igm})$, the cellwise quantity from SB simulation. For each snapshot $z_i$ we evaluate the fitting parameters $\rm a_i$, $\rm b_i$ and $\rm c_i$ using the coarse-grid field we derived in \S\ref{sec:CoarseGridMeth}. 

\subsubsection*{III) Stochastic Subgrid Clumping (SC) Model}
\begin{figure*}\vskip-4mm
	\includegraphics[scale=0.95]{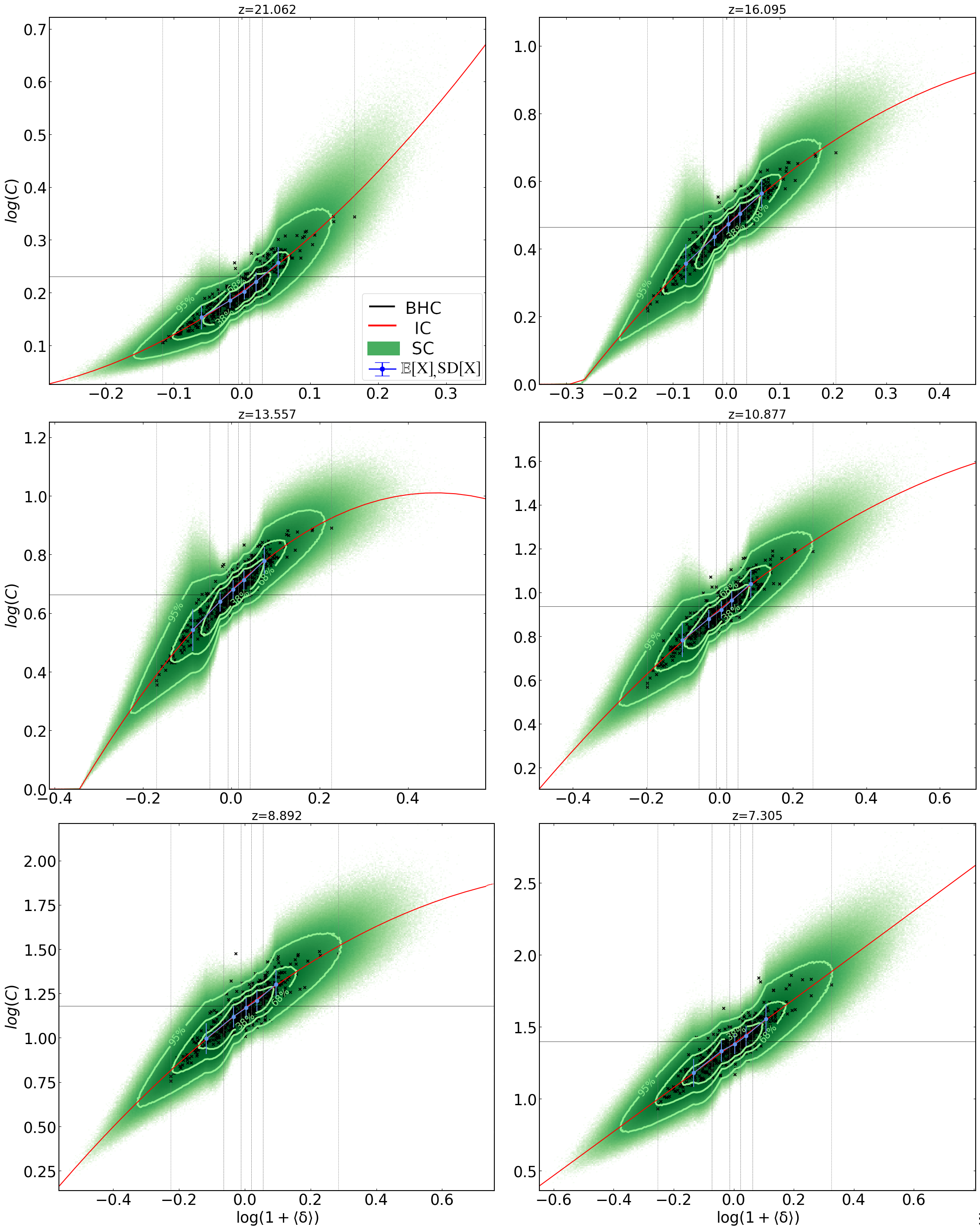}\vskip-2mm
	\caption{Realization of clumping factor for LB-2 at different redshift. The horizontal line (solid black) is the globally averaged clumping factor BHC. In red the one-to-one fit IC. Blue error barred points represent the expected value and standard deviation of the log-normal distribution. Vertical lines (grey dashed) indicate the bin limits. The green area indicates the SC realization estimated by \autoref{eq:realiz_ln}. We plot the $38\%$ ($\rm 0.5\,\sigma$), $68\%$ ($\rm 1\,\sigma$) and $95\%$ ($\rm 2\,\sigma$) confidence interval to highlight the realization distribution. Cross point are the \textit{coarse} SB data used to calibrate the model parameters. In the case of $\rm z=7.305$ they correspond to the one of \autoref{fig:corr+pdf} (left column).}
	\label{fig:plot500}
\end{figure*}
\label{SC_model}
This model, first presented here, aims to account for the natural stochasticity of the relation between local clumping and overdensity, as observed in full numerical simulations.
This stochasticity is due to various environmental effects beyond the dependence of the clumping on the local density, and results in a significant scatter around the mean relation used in the IC model (Fig.~\ref{fig:corr+pdf}).

We model this scatter from the simple one-to-one relation by binning the SB coarse-grained clumping in several (here five) wide bins of overdensity $\rm \Delta \delta_j$. In each bin we fit the scatter using a log-normal distribution.
\begin{equation} \label{eq:lognorm}
	\rm \mathcal{P}(x\,|\,z_i\,,\,\Delta \delta_j) \equiv \frac{1}{x\,\sigma_{ij}\,\sqrt{2\pi}} \exp{\left(-\frac{(\ln(x)-\mu_{ij})^2}{2\,\sigma_{ij}^2}\right)} 
\end{equation}
where $\rm x=C_{\igm}$. For each snapshot $\rm z_i$ and bin $\rm \Delta \delta_j$ we evaluate and record the parameters $\rm \mu_{ij}$ and $\rm \sigma_{ij}$.

A stochastic process is then applied to generate log-normal random values from two dimensional uniformly distributed variable $\rm u_1,\,u_2 \in [0,1]$, by using a modified\footnote{A random variable is defined log-normal distributed when the natural logarithm of the variable is normal distributed. Therefore our modification simply consist in taking the exponential of the transformation.} \textit{Box-M\"{u}ller transformation}.
\begin{equation}\label{eq:realiz_ln}
	\rm C_{SC}(z,\,\Delta \delta|\,\mu_{ij},\,\sigma_{ij}) = e^{\,\mu_{ij} + \sigma_{ij} \cdot \sqrt{-2\,ln(u_1)\cdot cos(2\pi\,u_2)}}
\end{equation}

where $\rm\mu_{ij}$ and $\rm\sigma_{ij}$ are the weighed log-normal parameters for LB-1 and LB-2.
Finally, we note that the range of overdensities in the SB simulation is inevitably narrower due to the smaller volume compared to our target reionization volumes. For data beyond the SB limits, for high and low densities, we fix the mean value to the one given by the IC model, while standard deviation is fixed to the one obtained in the closest density bin.

These distributions are then sampled randomly to create realisations of the clumping in large-volume simulations. A similar approach, but in a different context, has been used previously by \cite{Tomassetti2014} and \cite{Lupi2018}, motivated by observation of density distribution in giant molecular clouds.\\

In \autoref{fig:corr+pdf} we show examples of the resulting parametrization obtained from the three models at redshift $\rm z=7.305$, applied at the LB-1 and LB-2 RT resolutions. We show the coarse-grained N-body data, along with the BHC and IC models, as well as the mean, $\rm \mathbb{E}[X]=e^{\mu+\frac{1}{2} \sigma^2}$, and the standard deviation, $\rm SD[X] = e^{\mu+\frac{1}{2}\sigma^2} \sqrt{e^{\sigma^2}-1}$, of our proposed log-normal distribution of the stochasticity. On the side plot we show the coarse-data distribution (shadow histogram) and the resulting log-normal fit (solid line) with brief description of the density-bin limits and fitting parameters shown in the legend.

\begin{figure*}[t]
	\centering
	\makebox[\linewidth]{\includegraphics[scale=0.22]{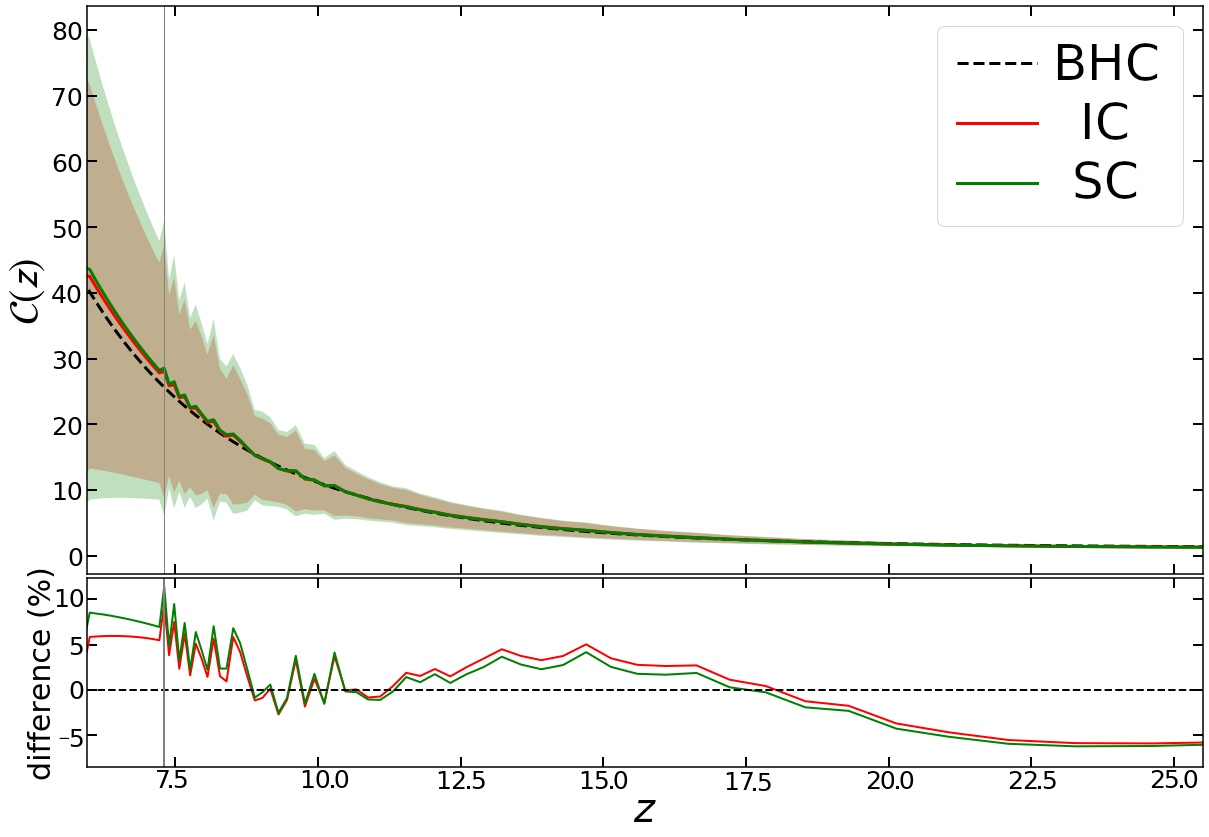}\includegraphics[scale=0.22]{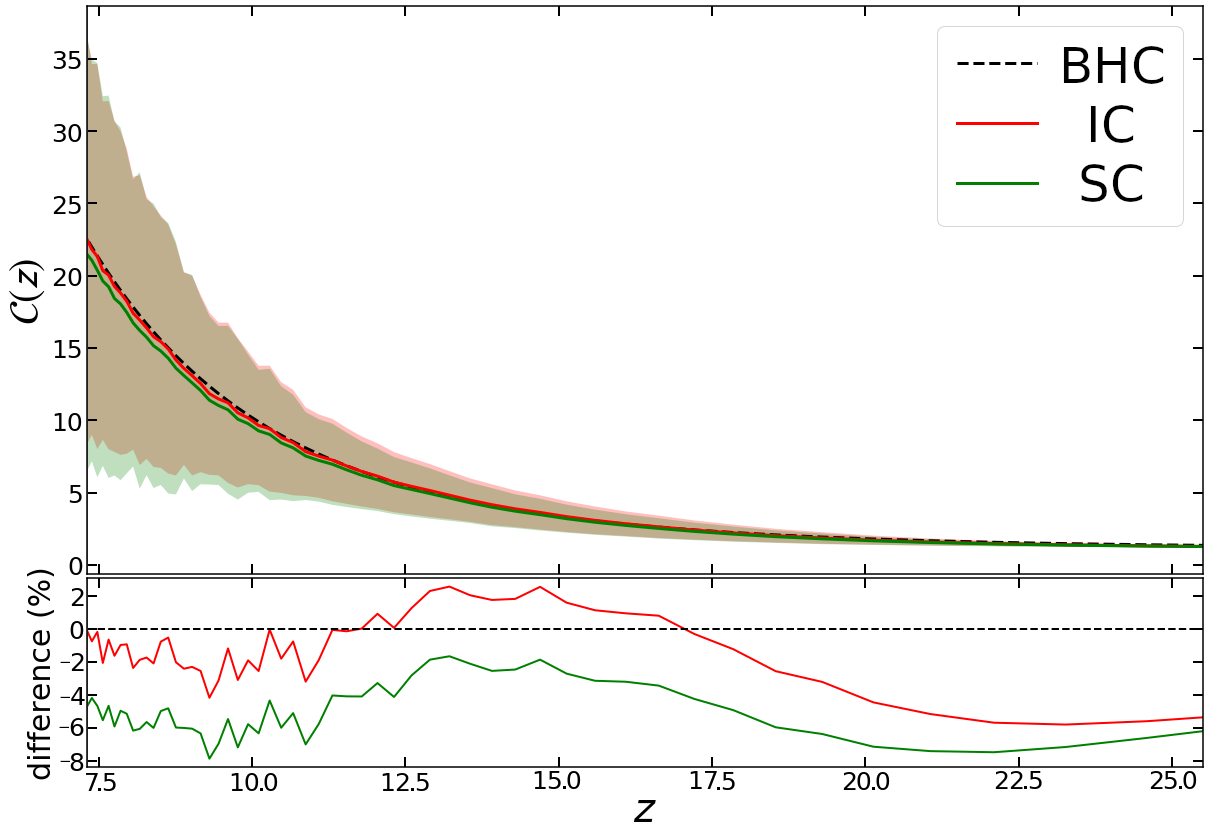} }\vskip-2mm
	\caption{Comparison of mean clumping values for the three different models, the redshift evolution of the mean clumping factor for the three models, respectively, shows the range of the standard deviation. On the bottom plot we show the relative where the left image is for LB-1 and right for LB-2. On the upper plot we have the dashed black line is BHC, in red IC and in green SC model, the shadow error in percentage of the difference with BHC.
	}\label{fig:mean_comparison}
\end{figure*}
\subsection{Clumping Implementation in Large-Scale Volumes} \label{sec:ClumpRealiz}
We used simulations LB-1 and LB-2 as examples of our method for creating large-volume
simulation sub-grid clumping realisation. Results are shown in \autoref{fig:plot500}.
In the figures we show the N-body data upon which the model is based (black crosses), the volume-averaged clumping factor BHC (black horizontal line), the one-to-one quadratic fit IC (red solid line), the expectation value $\rm \mathbb{E}[X]$ and the standard deviation $\rm SD[X]$ of the log-normal distribution in each density bin (blue error-bar points) with the relative bins limits also shown (dashed vertical line). Finally, our SC model clumping realisation (green area) based on the density field of LB-2 is shown with contours corresponding to the $95\%$ (outer), $68\%$ (middle) and $38\%$ (inner) confidence interval. Tables with parameters of the three models used in this paper can be found online\footnote{table for model parameters:\\ \url{https://github.com/micbia/SubgridClumping}}.

The results illustrate the extend to which each subgrid clumping model reproduces the trends in the direct N-body data throughout the evolution. The BHC (mean-clumping) model roughly matches
the peak of the contours and its evolution over time. The IC model (quadractic fit) captures well the general trend of the density-clumping relation and tracks well the highest density of
data points. Finally, our new SC model realisation fully reproduces the data, including the scatter around the mean relation. The contours trace the majority of the simulation data quite closely, apart from a few outliers. However, a few things should be noted here. 

First, as noted in \S~\ref{SC_model}, the large volumes generally sample much wider range of environments than smaller ones used to produce the model, thus inevitably the large-volume
realisation should extrapolate to over-densities outside the range sampled by the direct N-body data, for both larger and smaller over-densities. Second, again as discussed above, for statistical reasons we fixed the bin sizes so that they contain same number of data points, which inevitably results in quite uneven bin widths. These are very narrow near the peak density
of points and are quite wide for extreme values of the over-density. The combination of these
factors yields the 'flairing' of the realisaion (green) points at both large and small values
of the over-density, and thus possible minor discrepancies with what we find in simulations. However, only small fraction of the points are in these regions, as demonstrated by the density of points, and therefore it is unlikely this will affect the results in any significant way. Obviously, the IC model is potentially affected in a similar way, since the quadratic fit
is used beyond the range of the original data points.

As a consistency check, we compare the redshift evolution of the volume
average of the clumping realizations based on the SC and IC models vs. the actual global mean $\rm\mathcal{C}_{glob}$ based on the simulated data (\autoref{fig:mean_comparison}). The vertical line indicates the redshift at which the SB simulation was stopped, thus data beyond that is extrapolated.
The relative errors of the mean values (bottom panels) are in agreement within the $6-7\%$ for LB-2 (right) and within $\rm 10\%$ for LB-1 (left), throughout the relevant redshift range $\rm 6\leq z\leq30$. At the highest redshifts ($z>30$) the errors appear larger, however over that redshift range the density fluctuations are small and thus all clumping factors converge to $1$ and do not contribute to the recombination rate.\\
Hence, the local density inhomogeneity does not significantly affect the global averages; however, we expect that the local clumping factor plays a greater role in the recombination and ionization at small scales (e.g. on the H II region size distribution, \textit{ionized bubble} volume evolution, etc.). The proposed models are roughly consistent with results of previous papers \cite{Shapiro2018}, \cite{Iliev2007} and once our the RT-simulation are performed we expect to obtain similar confirmation from the work of \cite{Iliev2012}.\\

\section{Clumping Model Effect on Observational Signature} \label{chap:RTresult}

The sub-grid clumping model employed affects the local IGM recombination rates, which is then reflected in the derived observable signatures of reionization. In order to understand and try to quantify the importance of this choice, we perform three RT simulations where we fix the source production efficiencies of ionizing photons and vary solely the clumping model. At each time step the
precomputed N-body density fields are used to create a realisation of the corresponding gridded clumping factor, as described in \S\ref{sec:ClumpMod}. These clumping grids are then stored and provided as additional inputs
to full radiative transfer simulations with the \texttt{C$^2$-RAY} code \citep{methodpaper}. Specifically, the simulation used for this section is
LB-1.



The simulation redshifts span the range $z=40$ to $6$, for a total of $125$ snapshots. The corresponding aperture on the sky vary from $3.6$ to $\rm4.7\,deg$ per side, and covers the redshifted 21-cm frequency range from $26$ to $\rm45\,MHz$. The resolution evolves from $43.5''$ to $\rm56''$ in the spatial direction, and from $0.08$ to $\rm0.15\,MHz$ in frequency.

\subsection{Reionization History}\label{sec:IonizHistory}

\begin{figure}
	\hskip-7mm\includegraphics[scale=0.22]{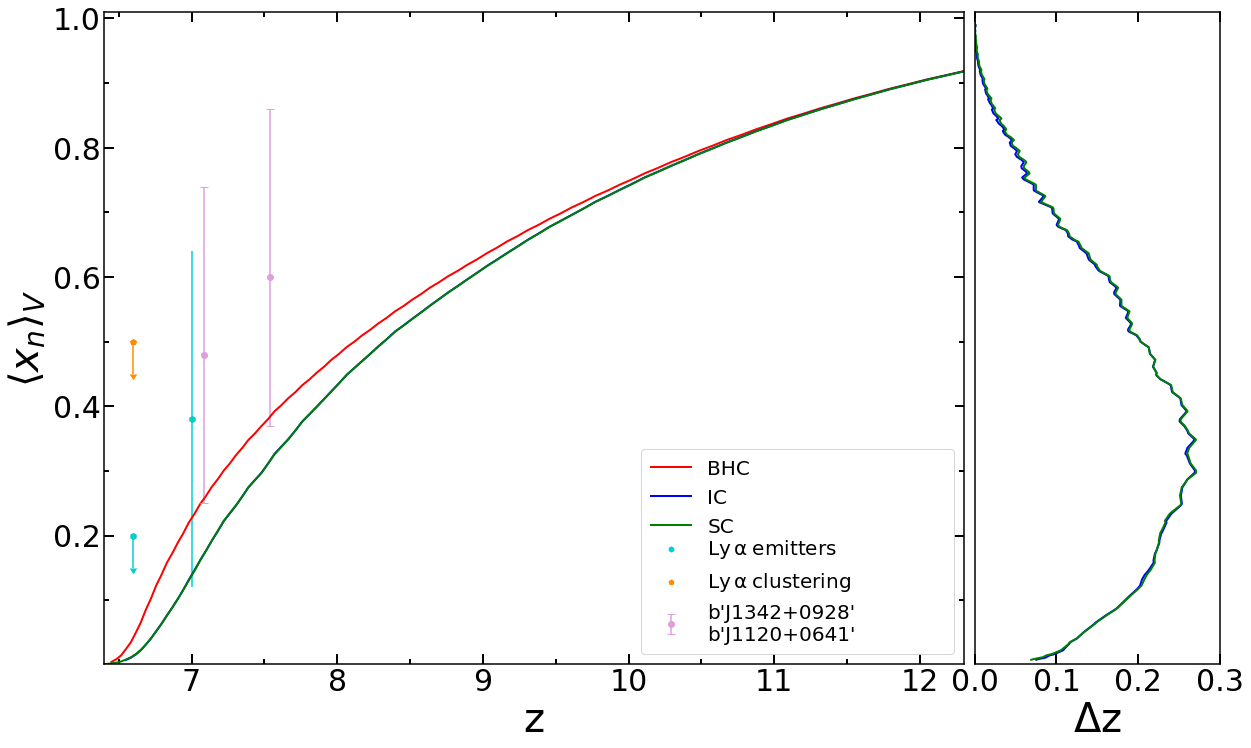} \vskip-2mm
	\caption{Left plot, the volume-averaged neutral fraction for BHC (solid red), IC (dashed blue) and SC (solid green) clumping models applied to simulation SB-2. On the right we show the redshift delay of IC and SC models compared to BHC. As a comparison we include observational constrains (see legend) from Ly$\alpha$ emitters (cyan circle) \protect\citep{Ota2008, Ouchi2010}, Ly$\alpha$ clustering (orange circle) \protect\citep{Ouchi2010} and from high redshift quasars spectra (pink) \protect\citep{Davies2018}.}
	\label{fig:xi} 
\end{figure}

Our results on the reionization histories are presented in Fig.~\ref{fig:xi}
and Table~\ref{tab:Eor_history}.
\begin{figure}
	\hspace{-3mm}\includegraphics[scale=0.32]{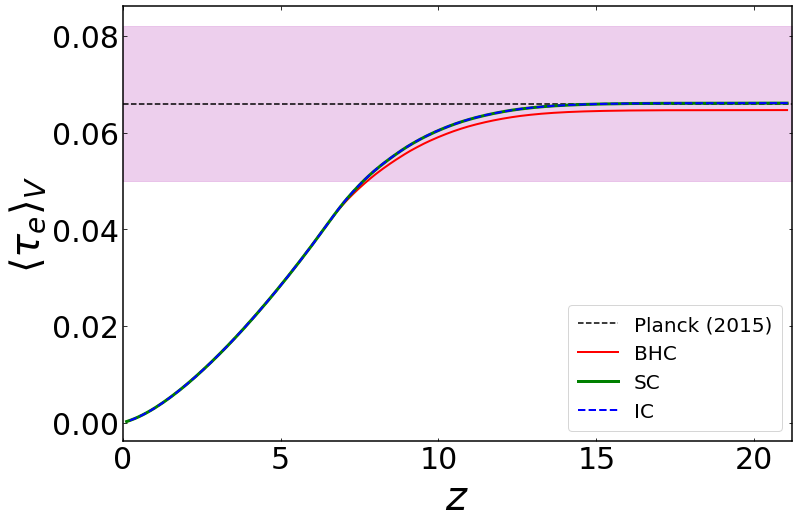}\vskip-3mm
	\caption{Thomson scattering optical depth to CMB photons integrated
		through our simulations, as labelled. Hown are also the
		\textit{Planck} observational constraint (black dashed line) along
		with its relative $1$-$\sigma$ confidence interval (violet shaded)} \label{fig:tau}
\end{figure}
\begin{table}
	\begin{tabular}{ccccccc} \hline
		Model				& $z_{10\%}$& $z_{30\%}$& $z_{50\%}$& $z_{70\%}$& $z_{90\%}$& $z_{reion}$	\\ \hline\hline
		BHC				& $11.918$	& $9.533$	& $8.118$	& $7.221$	& $6.721$	& $6.483$		\\
		IC				& $11.918$	& $9.611$	& $8.340$	& $7.480$	& $6.905$	& $6.583$		\\
		SC 				& $11.918$	& $9.611$	& $8.340$	& $7.480$	& $6.905$	& $6.549$		\\ \hline
		$\Delta z$			& $0$		& $0.078$	& $0.222$	& $0.259$	& $0.184$	& $0.1$			\\
		$\Delta t\,[Myr]$	& $0$		& $5.8$		& $22.9$	& $34.4$	& $28.9$	& $17.2$	\\ \hline
	\end{tabular}
	\caption{Mean volume-averaged ionized fractions, $\bar{x}_i$, at a reionization milestones: $10\%$, $30\%$, $50\%$, $70\%$ and $90\%$ volume of the gas ionized. The last column $z_{reion}$ lists the end of reionization, defined as $\bar{x}_i=99\%$. The second section lists the redshift and time differences with respect to the BSC model.}
	\label{tab:Eor_history}
\end{table}
\begin{figure*}
	\hskip-5mm\includegraphics[scale=0.27]{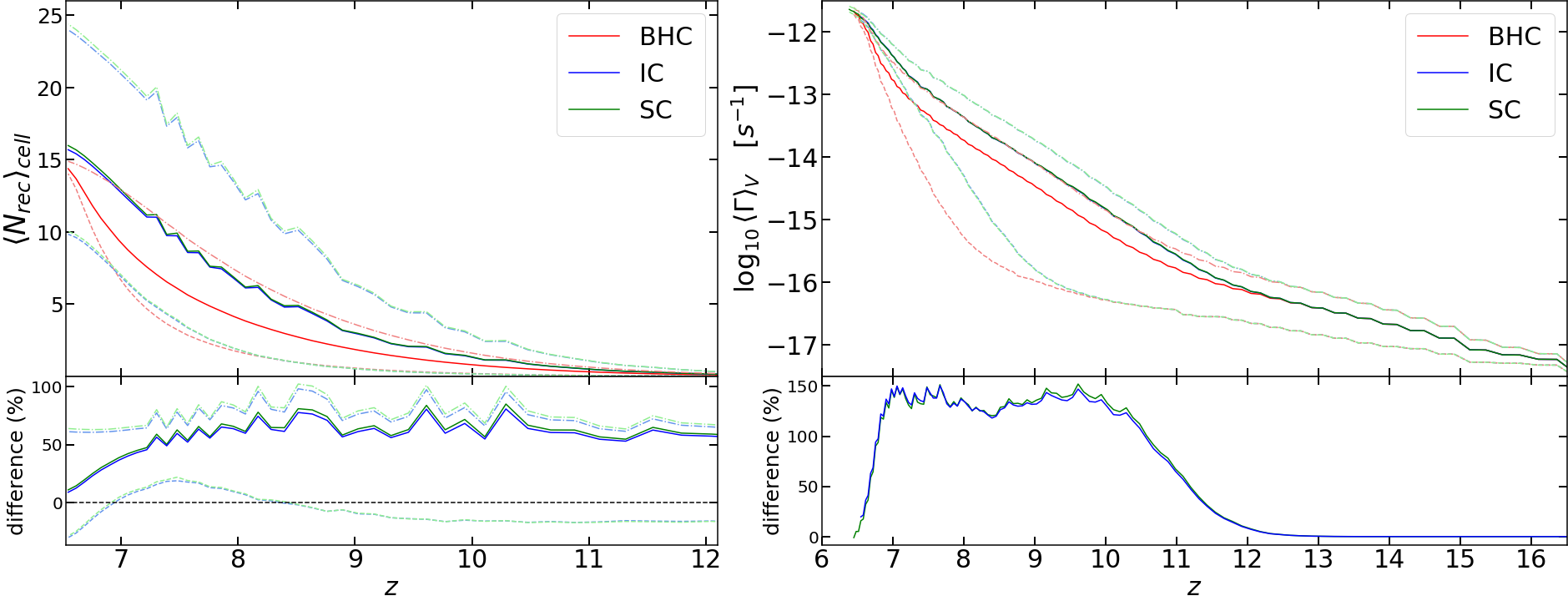} \vskip-2mm
	\caption{Evolution of the number of recombinations per hydrogen atom and per Hubble time throughout reionization (left) and the volume-averaged photoionization rate (right). The bottom plots show the relative difference compared to BHC in each case. Dashed and respectively dashed-dotted lines of the same colour indicate the relative quantity in under-dense and over-dense regions.
	}\label{fig:rec+phot} 
\end{figure*}
Perhaps counter-intuitively, either of the more realistic, density-dependent clumping treatments (SC and IC) yield somewhat faster evolution and an earlier end of reionization compared to the BHC model. The former models diverge from BHC around $z \leqslant 12$, and thereafter the mean reionization is accelerated with a maximum difference at $\bar{x}_i=70\%$, of $\Delta z \simeq0.3$ at $z\simeq7.5$, corresponding to a time difference of approximately $\rm36\,Myr$. The end of reionization is delayed by $\Delta z=0.1$, or $17$~Myr. Here there is very little difference between the SC and IC models. Compared to the observational constraints, all three models reionize somewhat early, however these constraints are largely upper limits, and with significant uncertainties. Moreover, our main interest is the relative effect of different sub-grid clumping models, rather than a faithful reproduction of the constraints.

During reionization, free electrons scatter CMB photons via inverse Compton scattering, suppressing CMB anisotropies on all scales and introducing polarization on large angular sizes. The contribution from free electron can be quantified by the integrated Thomson scattering optical depth along the line of sight, given by
\begin{equation}\label{eq:tau}
\tau_e(z) = c\,\sigma_T \int^z_{0} \frac{n_e(z')}{(1+z')\,H(z')} dz' 
\end{equation}
where $\sigma_T = 6.65\times 10^{-25} cm^2$ is the Thomson cross section, $c$ the speed of light and $n_e$ is the electron density at a given redshift.\\
In \autoref{fig:tau} we plot the volume mean of \autoref{eq:tau}, integrated back in redshift. In agreement with the global reionization histories, the inhomogeneity-dependent models are very similar to each other and are slightly optically thicker than the BHC case, due to the more advanced reionization in the latter. Regardless of this small difference, all three cases are in close agreement with the \textit{Planck-LFI} 2015 results \cite{PlanckCollaboration2015}, which found $\tau_e=0.066\pm0.016$ corresponding to an instantaneous reionization for redshift $z_{reion}= 8.8^{+1.7}_{-1.4}$.

The importance of recombinations throughout reionization could be quantified
by the (dimensionless) mean rate of recombinations per hydrogen atom per Hubble time:
\begin{equation}\label{eq:dimrec}
\rm\left< \dot{N}_{rec} \right> 
= \frac{\left<\mathcal{R}\right>}{t_H(z)\, \left<n_{\rm H}\right> \, V_{cell}} 
= 0.72\, \frac{\alpha_B\, (1+z)^3\,}{H(z)} \frac{\bar{\rho}_{c,0}\,\Omega_{b}}{\mu_H \, m_p}\, C_{\rm HII} \left< x_{\rm HII} \right>^2
\end{equation}



In \autoref{fig:rec+phot} (top left) we show the evolution of the mean of this quantity over the full simulation volume (solid lines), as well as averaged only over the over-dense (dashed lines) and under-dense (dot-dashed lines) regions. Colours indicate the model used, as per legend. We also show (bottom left) the relative percentage difference compared to the BHC model. As could be expected, the number of recombinations grows strongly over time, starting close to zero,
departing from BHC model around $z\sim12$, and then all reaching $\sim 15$ at late times, as more and more structures form over time. Although all models end up at similar values by $z\sim6$, the BHC model lags behind throughout the evolution. The IC and SC models yield very similar values at all times. The over-/under-dense volumes yield much higher/lower number of recombinations, respectively, demonstrating the wide variety of outcomes
dependent on the local conditions. Interestingly, the over-dense average for the BHC model results in very similar recombinations to the full-volume averages of SC and IC models, showing that at least on average the clumping in these last models behaves the same way as the over-dense regions in BHC. Overall, the SC model shows a few percent higher recombination rate ($\sim1-5\%$) compared to the IC model. This is most likely due to the stochastic nature of the realization process, also related to the broader scatter in \autoref{fig:mean_comparison} (shaded areas).

In \autoref{fig:rec+phot} (right panels) we compare the (non-equilibrium) photoionization rates $\Gamma_i$ computed during the run by the \texttt{C$^2$-Ray code}. Just as above, all mean photoionization rates are
essentially the same until $z\sim12$, after which the BHC model one rises more slowly, lagging behind the other two cases by about factor of 2.5 throughout
most of the evolution, eventially catching up by $z\sim6$. The average rates
in the overdense regions are higher than the mean (reflecting the inside-out nature of reionization) by a similar amount, while
the mean photoionization rates in the under-dense regions lag behind by larger
factors, up to several hundred, before again rising steeply and catching up with
the mean by $z\sim6$. Interestingly, the mean rate in BHC overdense regions is
again very close to the whole volume means of IC and SC models. The average
values in the under-dense regions remain the same for all models until much
later, $z\sim9.5$, indicating that the specific clumping model has little
influence before that redshift.

\begin{figure}
	\hskip-5mm\includegraphics[scale=0.26]{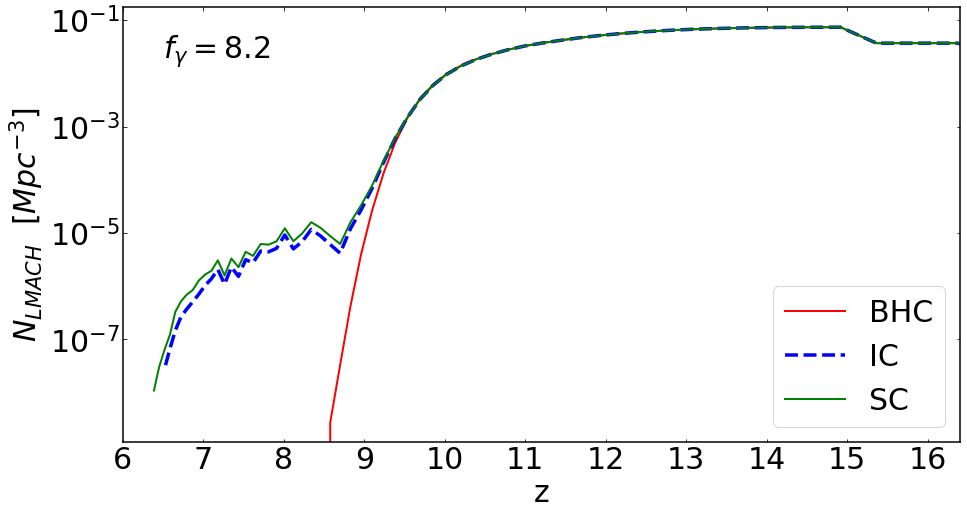} \vskip-2mm
	\caption{The number density evolution of unsuppressed LMACHs ($f\rm_{\gamma}=8.2$). Solid red line the BHC model, dashed blue line the IC model and in gree the SC model.} \label{fig:tot_LMACH} \vskip-2mm
\end{figure}

\begin{figure}
	\hskip-3mm\includegraphics[scale=0.27]{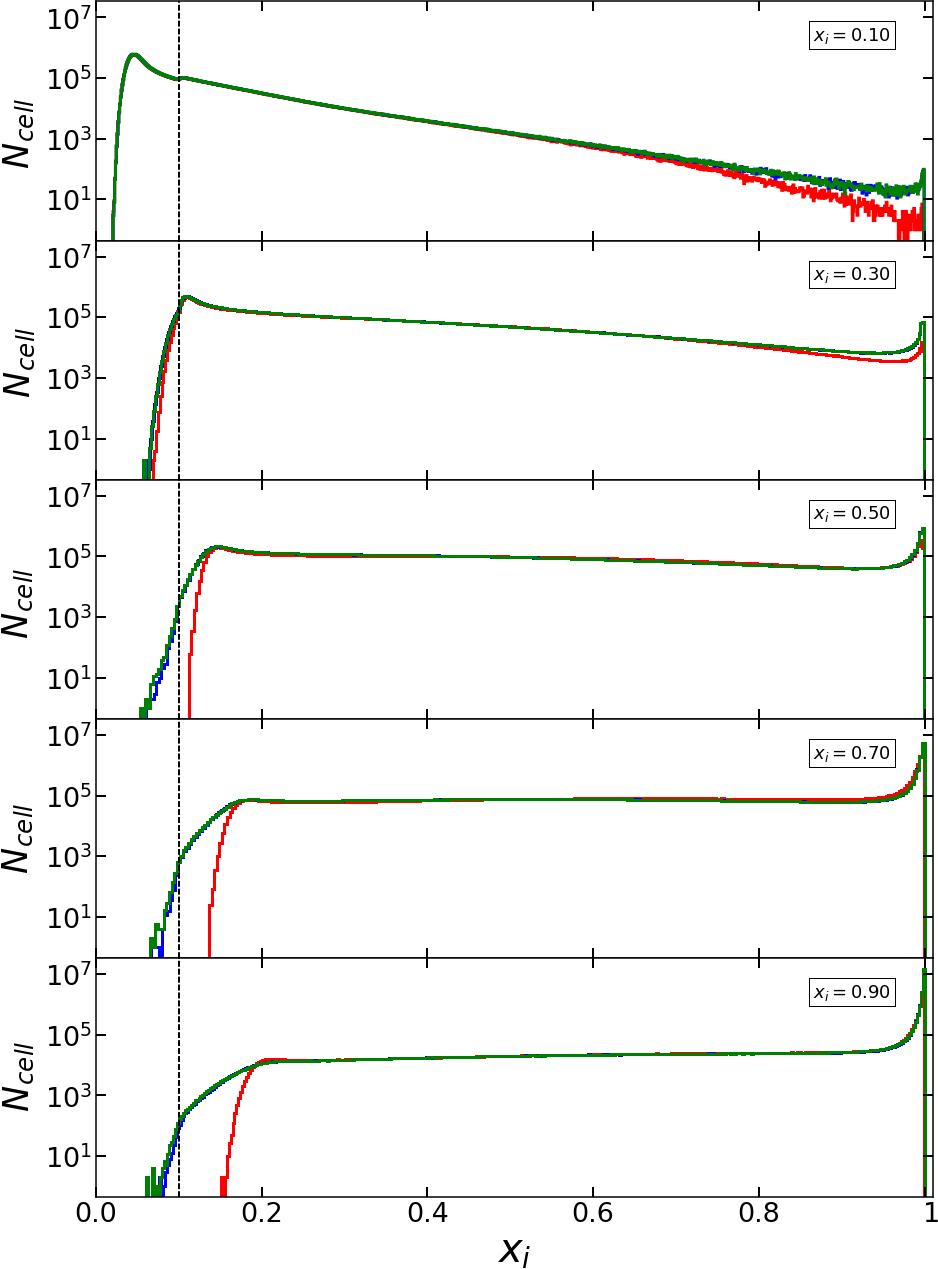}\vskip-2mm
	\caption{Ionized cell number density at reionization milestone $\rm \bar{x}_i=0.1,\,0.3,\,0.5,\,0.7,\,0.9$, top to bottom, for model BHC (red), IC (blue) and SC (green). Vertical line (black dashed) indicates the ionization threshold, $x_{\rm i}=0.1$, for partial suppression of LMACHs.}
	\label{fig:xi_PDF}
\end{figure}

At first glance, it seems somewhat counter-intuitive that reionization proceeds faster in the denity-dependent models IC and SC, despite their notably higher recombination rates. The reason for this is that in the former cases also the suppression of low-mass galaxies (LMACHs) due to radiative feedback is weaker than in the BHC case, as illustrated in \autoref{fig:tot_LMACH}. In the BHC case essentially all such galaxies are suppressed by $z\sim8.5$, while in the density-dependent models the suppression is slowed down, allowing LMACHs to last longer in high density regions. This is further clarified in \autoref{fig:xi_PDF} where we show the number density distribution of ionized fraction of cells at five different reionization stages, $\bar{x}_i=0.1,\,0.3,\,0.5,\,0.7,\,0.9$, approximately corresponding to redshift between $z\simeq 12-6$ (see \autoref{tab:Eor_history}). The vertical line indicates the partial suppression threshold for LMACHs. Early on ($\bar{x}_i=0.1$) the gas clumping has yet had very litte effect, due to the still small ionized fraction and the short time available for recombinations, thus all models yield very similar results, with only BHC showing slightly fewer highly ionized cells. As reionization progresses ($\bar{x}_i=0.3$), IC and SC models remain very similar, while BHC is gaining more ionized cells, and at the same time it is starting to show a lack of neutral regions. Starting from roughly mid-point of reionization ($\bar{x}_i=0.5$), the dearth of neutral cells becomes ever more prominent whereas the peak of highly ionized cells stays roughly similar for all models.
A faint difference between SC and IC is visible at late times, where slightly more cells remain neutral in SC.

\subsection{Reionization morphology} \label{sec:SliceComparison}
\begin{figure*}
	\vskip-3mm\makebox[\linewidth][c]{\includegraphics[scale=0.6]{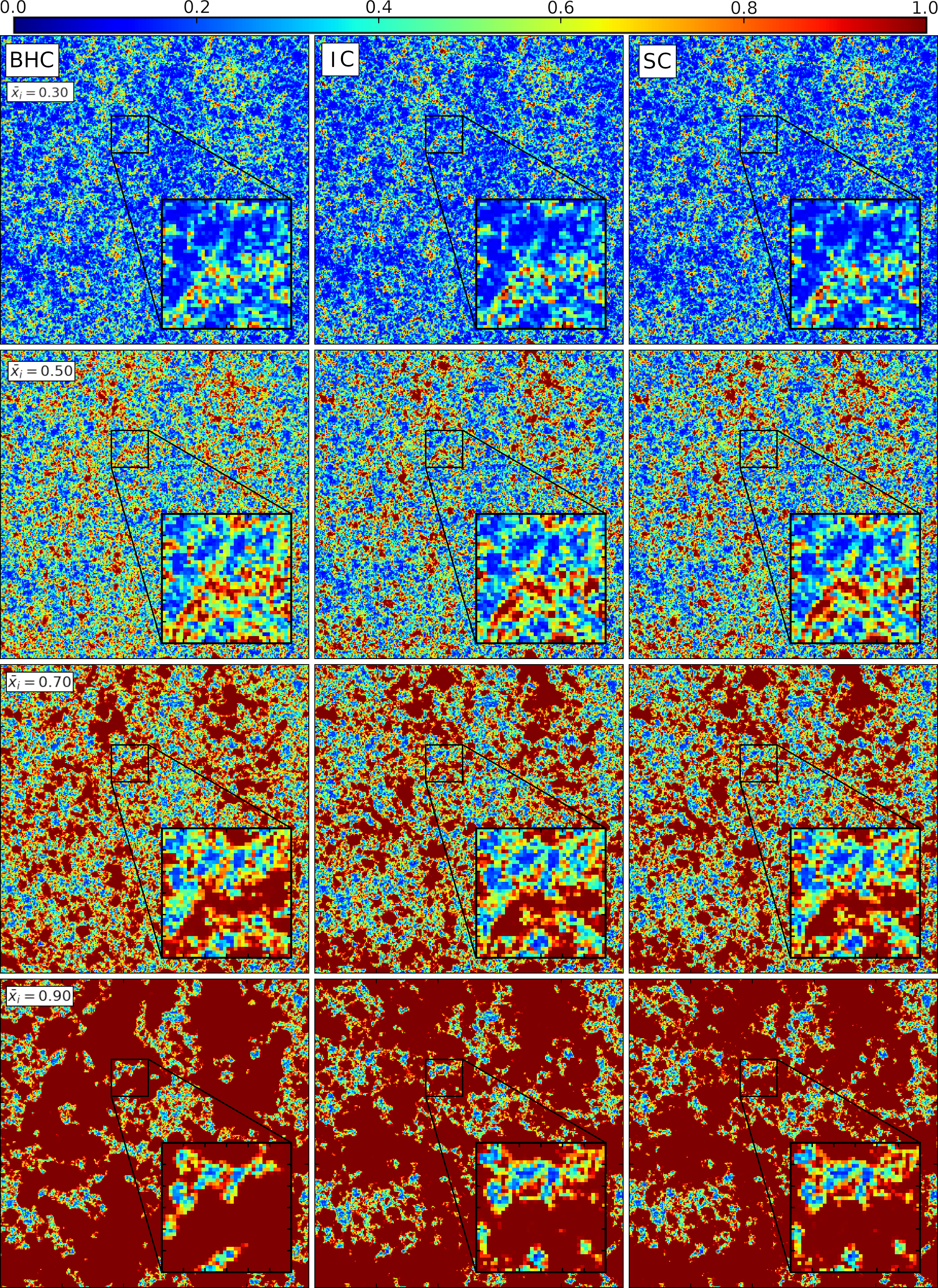}}\vskip-2mm
	\caption{Box slice comparison of LB-2 ionization fraction for different clumping models. In red/crimson highly ionized regions $x_{\rm i}>0.9$, in green/aquamarine transition $x_{\rm i}\approx0.5$ and in dark blue neutral regions $x_{\rm i}<0.1$. The zoom-in covers an area of $\rm85\mpc$ per side and each pixel represent a volume element of $\rm2.381\mpc$ per side. We compare slices at same global average ionization fraction, from top to bottom row we have $\rm\bar{x}_i=0.3,\,0.5,\,0.7,\,0.9$ (see \autoref{tab:Eor_history} for corresponding redshifts). From left to right column respectively we show the models BHC, IC and SC.}
	\label{fig:slice_comparison} 
\end{figure*}

The globally-averaged quantities discussed above (\autoref{fig:xi}, \ref{fig:tau} and \ref{fig:rec+phot}) give an overall idea of the reionization history. Next step is to understand how the sub-grid gas clumping model affects the propagation of radiation and the local features of reionization. In \autoref{fig:slice_comparison} we show box slice of LB-2 and compare simulation snapshots with similar globally averaged ionized fraction and the three gas
clumping models. From top to bottom row we have $\bar{x}_i=0.3,\,0.5,\,0.7,\,0.9$ (in \autoref{tab:Eor_history} we list the corresponding redshift at which this occurs and its consequent time delay
compared to the BHC model) and from left to right column we have the different models BHC, IC and SC. Red/crimson regions indicate highly ionized cells $x_{\rm i}>0.9$, in dark blue neutral regions $x_{\rm i}<0.1$, and in green/aquamarine the transition phase $x_{\rm i}\approx0.5$. Within each image we embed a zoom-in region, of $85\mpc$ per side, to better appreciate the morphological changes of a randomly selected under-dense neutral clump, as ionized fronts expand (bluer blob, right column plots).\\

Our simulations reproduce the general reionization features found in previous simulations \cite[e.g.][]{Iliev2014, Hutter2021astraeus}. In high density regions LMACH are the first halos to form. In our simulations they make their first appearance at redshift $z=21$, and by $z\sim12$ every volume element contains at least one ionizing source. At first, a modest number of isolated sources, highly clustered on small scale but homogeneously distributed on large scale, start to ionize their surrounding gas, forming small regions of a few Mpc size. The presence of sub-grid gas clumping slows down the propagation of the I-fronts and yields somewhat smaller, more fragmented H~II regions. Throughout reionization, these HII bubbles grow and eventually overlap, at which point the ionization process accelerates and many of the smaller bubbles percolate to much larger connected volumes. 

The side-by-side comparison shows some notable differences between BHC and the two density-dependent models, with the latter starting at a faster pace, with
earlier local percolation, then slowing down compared to the former case. Modest differences appear between the three models in terms of large scale morphology, with a higher degree of ioniziation around early sources in the density-dependent models IC and SC (respectively central and right panel).
From around the mid-point of reionization ($50\%$ ionization by volume, second row of images) we can see neighbouring growing regions connecting to each other and starting to highly ionize the linking filament. At this point, accordingly to \autoref{fig:xi_PDF}, all cells in BHC have surpassed the threshold limit $x_{\rm i}=0.1$ for the partial suppression of low mass haloes. For IC (middle) and SC (right column) the degree of ionization around sources is visibly more intense compared to BHC, in fact we can distinguish highly ionized cells clustered around the high density peak, whereas under-dense regions are kept fairly neutral. This diversity is due to the higher recombination rate in inhomogeneity dependent model, shown in \autoref{fig:rec+phot} (left), that effectively reduces the number of photons able to escape the cells of origin and spread into the neighbour grid elements. This is not the case for BHC, to which clumping factor in high density regions is underestimated and ionizing photons are free to percolate and been absorbed elsewhere in the surrounding IGM, therefore interconnecting filament cells between sources clearly appears extended and in a more advanced neutral-ionized transition (blue/aquamarine).

Later on ($x_{\rm i}=0.7$, third row of images), ionized regions have grown substantially and become strongly ionized. A first look suggests similar structure patchiness on large scale, although from the zoom-in we can observe that BHC has a wider and smoother transition between the ionized/neutral phases, whereas IC and SC show a narrower front, allowing more cells that host under-density to stay neutral. When the same transition region dwell across the three model, density dependent model show more irregularity with occasionally one or few cells appearing slightly more ionized then their surrounding.

The morphology differences are more evident at late times ($x_{\rm i}=90\%$, bottom row of images), whereby HII bubbles connect together to form one vast interconnected highly ionized region. At this stage the vast ionized IGM in IC and SC show variations that follow the higher recombinations due to density fluctuations, which is not the case in BHC model and therefore the same regions appear uniformly highly ionized, $x\approx 1$. On the other hand there are no striking difference between IC and SC, except for small variations, of a few pixels of size, on the ionized/neutral boundaries. We suspect that this is numeric artefact due to the stochastic nature of SC. We are developing a more complete clumping model, that we will present in future work, to exclude this uncertainty.

\subsection{Bubble Size Distribution} \label{sec:BubbleSize}

\begin{figure*}
	\includegraphics[scale=0.33]{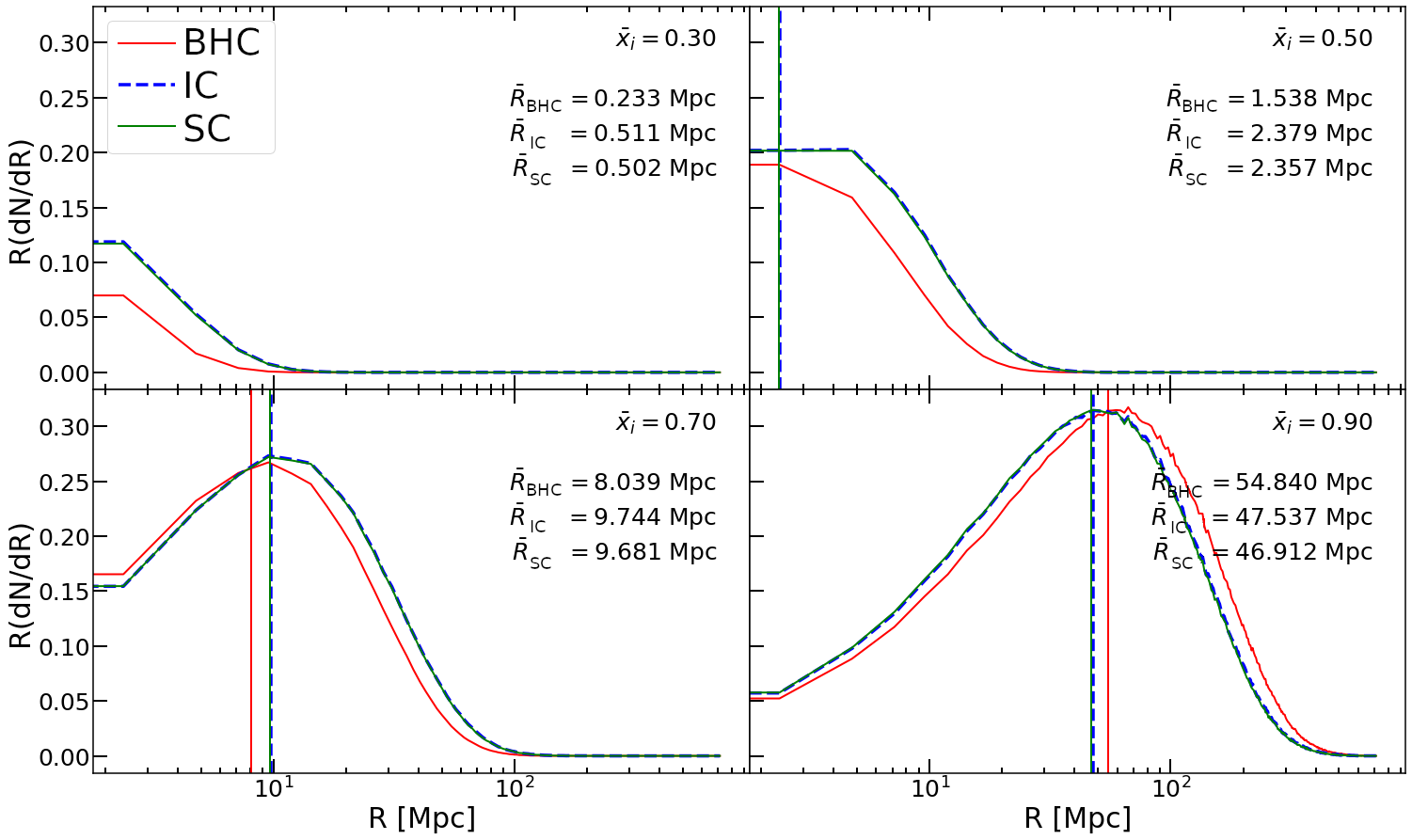} \vskip-2mm
	\caption{Ionized bubble size distribution for simulation LB-2 and
          the three gas clumping models BHC (red, solid), IC (blue, dashed)
          and SC (green, solid) at volume averaged ionized fractions $\rm\bar{x}_i=0.3,\,0.5,\,0.7,\,0.9$, as labelled. Vertical lines indicate the mean bubble radius $\rm\bar{R} =\int (R\,dN/dR)dR$ for the respective models.}
	\label{fig:bubble}
\end{figure*}
\begin{figure*}
	\includegraphics[scale=0.33]{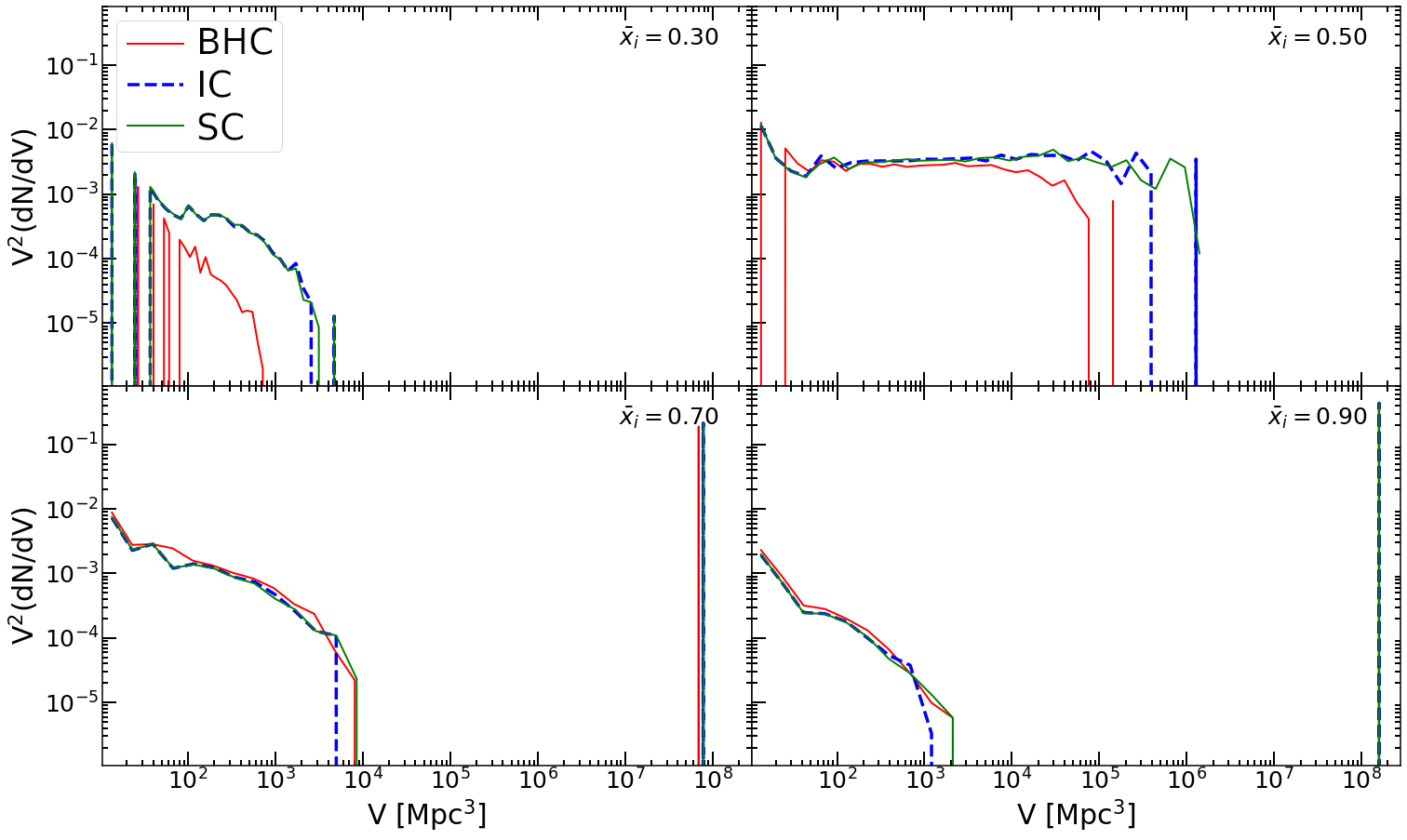} \vskip-3mm
	\caption{Ionized volume size distribution for LB-2. In red the result for BHC, in blue for IC and in green for SC. Distribution represents stages where the volume averaged ionized fraction is $\bar{x}_i=0.3,\,0.5,\,0.7,\,0.9$.}
	\label{fig:volumes}
\end{figure*}

One of the key characteristics of reionization, which directly affects all
observables is the normalized distribution of bubble sizes $R\,dN/dR$ or volume sizes $V^2\,dN/dV$ of ionized regions \citep{Furlanetto2004}. A number of complementary approaches to calculate these distributions have been proposed \citep[e.g.][]{Friedrich2011, 2016MNRAS.461.3361L,Giri2018}. Here we employ the Mean-Free-Path (MFP) method to calculate $R\,dN/dR$, and the Friends-of-Friends (FOF) algorithm \citep{2006MNRAS.369.1625I} to obtain $V^2\,dN/dV$ bubble size distributions (BSD). For both methods we employ the \texttt{TOOLS21CM}\footnote{\url{https://github.com/sambit-giri/tools21cm}} python package for EoR simulations analysis \citep{Giri2020}. In both cases, we apply a threshold value of $x_{th}=0.9$, since we want to highlight differences in distribution of highly ionized regions that develop around sources.

Results are shown in \autoref{fig:bubble} and \autoref{fig:volumes}, respectively we see the typical traits of the percolation process, with volume ranges that roughly corresponding to what is expected from large simulated box \citep{Iliev2014}. We present our results at four different reionization milestones, $\bar{x}_i=0.3,\,0.5,\,0.7,\,0.9$, see \autoref{tab:Eor_history} for corresponding redshifts. In the case of MFP-BSD, we calculate the mean bubble size by $\bar{R} =\int (R\,dN/dR)\,dR$, represented by the corresponding vertical lines for each simulation. The sharp cut-off at small scales $2.381\mpc$, for MFP-BSD, and $13.498\mpc^3$ for FOF-BSD correspond to the simulation cells size and volume respectively.

Early-on ($\rm \bar{x}_i=0.3$, top left panel, \autoref{fig:bubble}), LB-2 hosts small H~II bubbles with radius smaller then $\rm 10\mpc$. For inhomogeneity-dependent models IC and SC, distributions present many more highly-ionized regions, indication of a faster radiation propagation around sources. All three distributions peak at the size corresponding to one cell. The same trend is confirmed by the topologically-connected FOF volumes (\autoref{fig:volumes}), which are however typically larger than MFP, with volumes between $\rm30-700\,\mpc^3$ for BHC and a wider distribution for IC and SC, from one cell up to a few thousand $\rm Mpc^3$.

Even though the number of bubbles increase as reionization progress, at $\rm \bar{x}_i=0.5$ (top right), the MFP-BSD remain similar. However, the FOF-BSD shows a qualitative transition when the small H~II regions start to percolate into much larger, connected one. Their sizes vary widely, with a broad flat distribution (plateau) at smaller scales ($\rm V<10^5-10^6\mpc^3$). However, BHC and IC also show a bifurcated distribution, with a second peak at large scales, at $\rm 10^5 \mpc^3$ for BHC and $\rm 10^6 \mpc^3$ for IC, indicating that percolation process has started \citep{Iliev2008,Iliev2014,Furlanetto2015}. Compared to BHC, the IC distribution is shifted toward larger sizes, such that the limit for the plateau and the percolation cluster are up to one order of magnitude higher. A narrower separation between these two volume range indicates that the merging of ionized region in BHC has just started \citep{Iliev2014, Furlanetto2015, Giri2018}, whereas in the case of IC this process is already ongoing. On the other hand, IC and SC distribution show similarity at small scale but they differ for larger volumes. The former distribution shows a constant and continuous range of scales from large volumes $\rm V\sim10^6\mpc^3$ down to one cell, sign that ionized regions are in principle less interconnected and therefore the presence of one dominant super cluster has not yet occurred.

During the later stages of the reionization process ($\rm\bar{x}_i=0.7$, bottom left) this
bifurcation of the FOF-BSD continues and strenghtens, with ever more small patches merging
into the large one, while smaller patches become fewer and on average ever smaller. At this stage the three models present similar volume distributions, whereas their MFP-BSD varies. BHC distribution starts to show a clear characteristic size peak. Albeit of similar shape, the BHC size distribution is clearly shifted to smaller scales, with the average
bubble size smaller by a few Mpc and the distribution peak at scale about a factor of 2 smaller (8 vs 15 Mpc).

Towards late reionization ($\rm\bar{x}_i=0.9$, bottom right), the volume limit for isolated regions to grow before merging is further reduced to $\rm V\sim10^3\,\mpc^3$, while the percolation cluster surpass volumes of $\rm 10^8\mpc^3$ (i.e. close to the full simulation volume) in all the three cases. In \autoref{fig:bubble}, the sizes distribution in the BHC model has surpassed the other two, with average radius of $\rm 54.84\mpc$. IC and SC show again similar distribution but with an increasing, although still minor, difference in the mean radius. Volume distribution in \autoref{fig:volumes} present a similar situation, the only difference between IC and SC consists in the value of the volume merging limit, with a difference up to $\rm1\mpc^3$.


\begin{figure*}
	\includegraphics[scale=0.52]{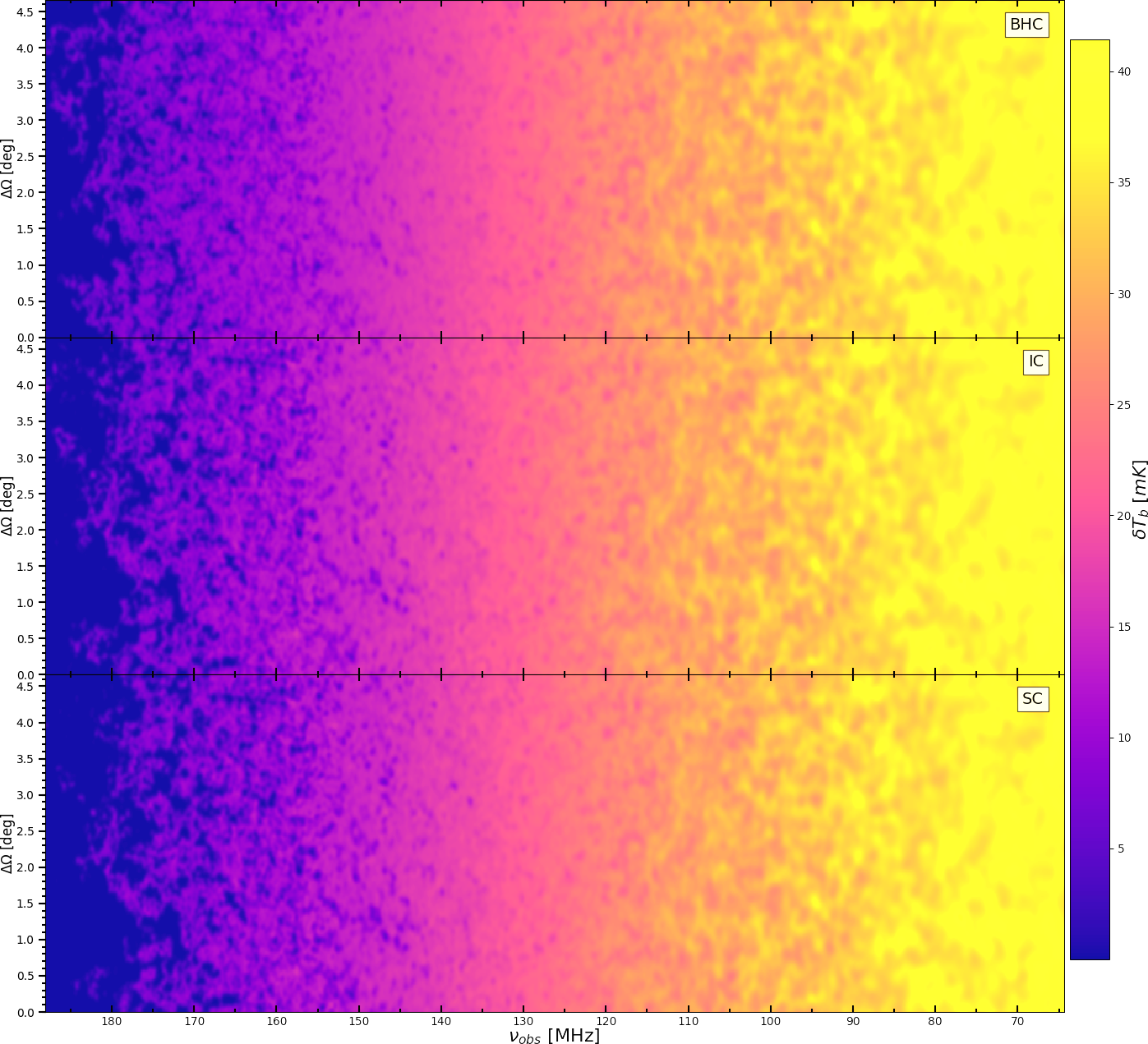}\vspace{-5mm}
	\caption{Smoothed differential brightness temperature lightcones, the colour map that shows the smoothed differential brightness $\rm\Delta T_b$ intensity as a function of redshifted \textit{21 cm} signal frequency $\rm\nu_{obs}$ and aperture $\rm\Delta \Omega$. The angular smoothing is performed by a Gaussian Kernel with FWHM $\rm\Delta \theta$, on frequency direction is done by a top-hat kernel with same width, we use a baseline of $\rm B=2\,km$ (maximum baseline of the core of SKA1-Low). The figure shows slice through the simulation and a comparison between BHC (top), IC (middle) and SC (bottom).}
	\label{fig:PL_dTb} 
\end{figure*}

\begin{figure}
	\includegraphics[width=\columnwidth]{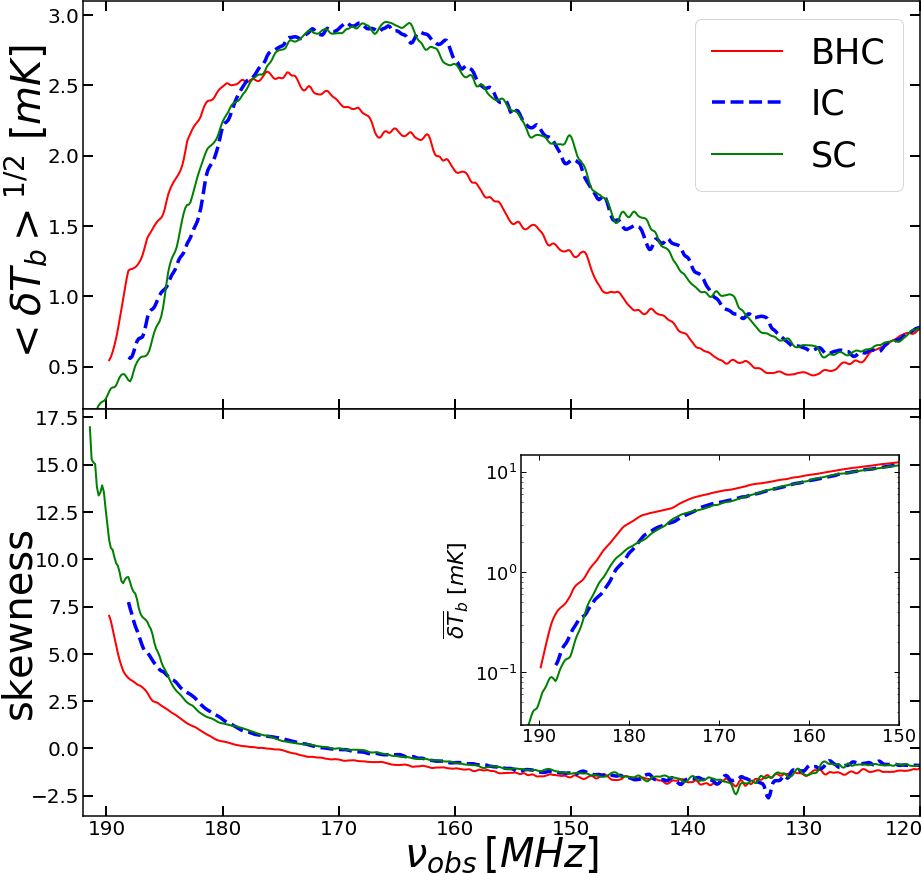}\vspace{-2mm}
	\caption{Differential brightness statistic quantities derived from the lightcones data smoothed on the core baseline of SKA1-Low ($\rm B=2\,km$). Plot on top shows the frequency evolution of the signal root mean squared (RMS). Bottom plot shows the skewness and an inset panel show the frequency evolution of the averaged differential brightness in logarithmic scale.}
	\label{fig:dTb_rms_and_mean}
\end{figure}

\begin{figure}
	\includegraphics[width=\columnwidth]{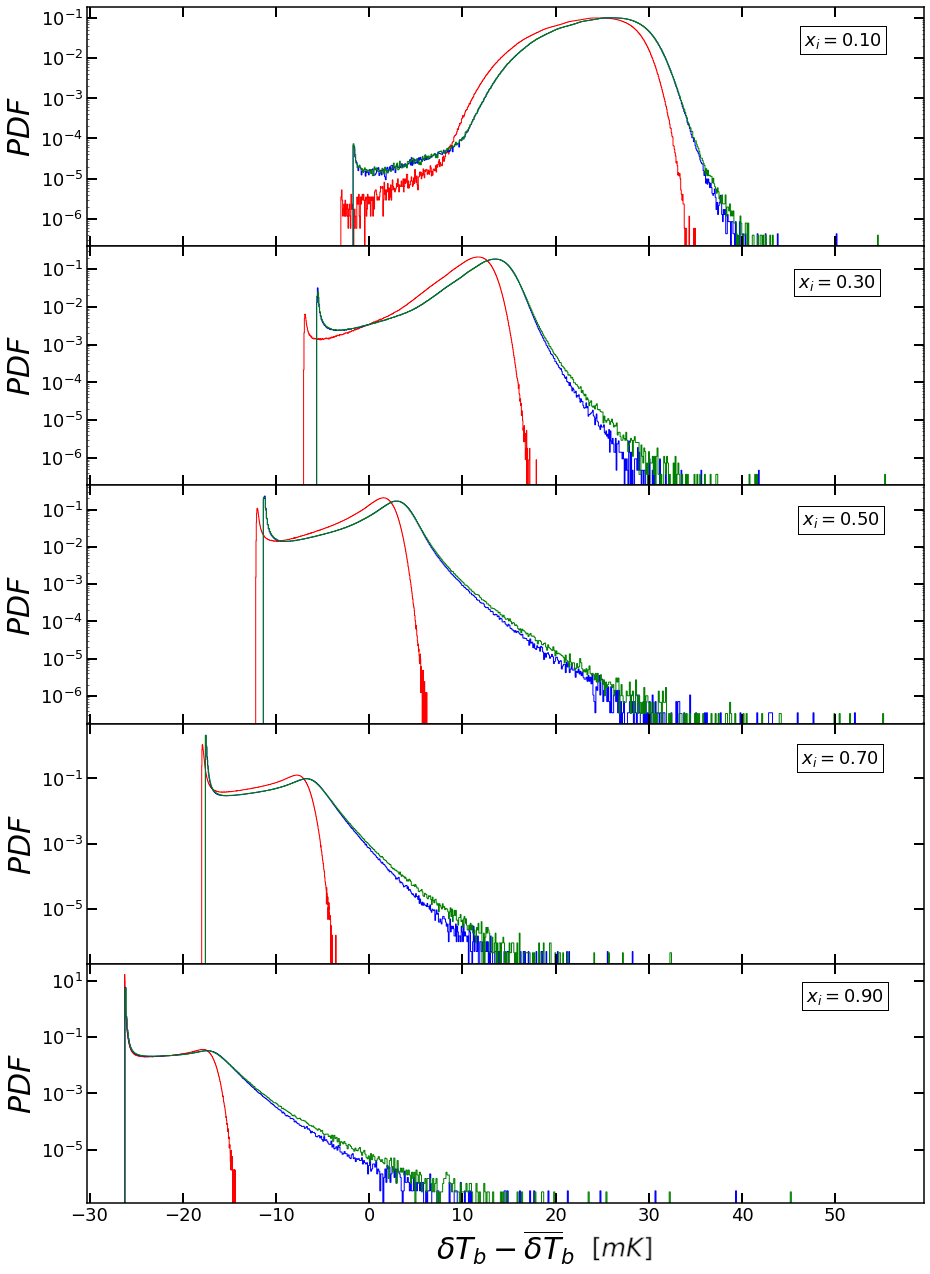}\vspace{-2mm}
	\caption{Probability distribution functions of the differential brightness temperature at ionized fractions $x_{\rm i}=0.1, 0.3, 0.5, 0.7$ and 0.9, for the three clumping models, as labelled.}
	\label{fig:21cm_PDF}
\end{figure}

\subsection{21-cm Signal Statistics and Power Spectra} \label{sec:21cmSign}
The hyperfine transition of neutral hydrogen redshifed into meter wavelengths
is a key observable of reionization. Its characteristic emission/absorption line has rest-frame wavelength $\rm\lambda_0=21.1\,cm$ and corresponding frequency 1.42 GHz. Radio interferometry telescopes measure the intensity of this signal by quantifying the differential brightness temperature $\rm\delta T_b\equiv T_b-T_{CMB}$ signal from patches of the sky, given as:
\begin{equation} \label{eq:dTb}
	\rm \delta T_b \approx 28\,mK (1+\delta)x_{HI} \left(1-\frac{T_{CMB}}{T_S}\right)\left(\frac{\Omega_b h^2}{0.0223}\right)\sqrt{\left(\frac{1+z}{10}\right)\left(\frac{0.24}{\Omega_m}\right)} 
\end{equation}
here $x_{\rm HI}$ is the fraction of neutral hydrogen and $\rm1+\delta = \left<n_{\rm N,IGM}\right>/\bar{n}_{\rm N,IGM}$ is the local IGM overdensity. The differential brightness is characterized by the relation between the CMB temperature $\rm T_{CMB}$ and spin temperature $\rm T_S$ (see e.g. \citealt{Furlanetto2006} and \citealt{Zaroubi2012} for extended discussion). \autoref{eq:dTb} saturates when the neutral hydrogen decouples from CMB photons and couples with the IGM gas heated by X-ray sources \citep[e.g.][]{Ross2019}, so that $\rm T_{S} \gg T_{CMB}$, which is the approximation we adopt here. This is known as the heating-saturated approximation where the signal is for the majority observable in emission, $\rm\delta T_b > 0$, true only at low redshift $z<15$. Thus in our simulation the approximated differential brightness is dependent on the density distribution of the neutral gas and redshift, such that $\rm \delta T_b \propto \sqrt{1+z}\,(1+\delta)\,x_{HI}$.

From the RT and N-body simulation outputs we calculate the differential brightness coeval cube at each time step. The cube is then smoothed in the angular direction by a Gaussian kernel with a FWHM of $\rm\lambda_0\,(1+z)/B$, where $\rm B=2\,km$ corresponds to the maximum baseline of SKA1-Low core. Smoothing along the frequency axis is done by a top-hat kernel with the same width and the above Gaussian kernel. SKA1-Low will not observe the coeval cube. Instead it will observe a lightcone, in which the signal evolves along the line of sight direction. We construct lightcones from our simulation results using the method described in \citet{Giri2018}. This method is also implemented in \texttt{TOOLS21CM}.

In \autoref{fig:PL_dTb} we show the smoothed lightcone for the three different clumping models, BHC, IC and SC, respectively from top to bottom. This type of data maps the 21-cm differential brightness evolution at the observed frequency $\rm\nu_{obs}=\nu_0/(1+z)$, where $\rm\nu_0=1.42\,GHz$ is the rest frame frequency when the signal was emitted at redshift $\rm z$. We then express the comoving box length in corresponding angular aperture of $\rm4.65^{\circ}$ at $\rm z=6.583$. 

Early on, the IGM remains mostly neutral, the average signal largely follows
Eq.~\ref{eq:dTb} ($\rm\delta T_b>30\, mK$) and the fluctuations are driven by the density distribution. The gas clumping also remains low and therefore at low frequencies, $\rm\nu_{obs}>120\,MHz$, there is no visible difference between simulations. As radiation escapes the host halos, it starts to form small isolated transparent regions around sources and gradually suppresses the average signal. The H~II regions are still small and thus are smoothed over by the observation beam. \autoref{fig:PL_dTb} shows very similar evolution for the three simulations at frequency higher then 130 MHz ($z<10$), but with different intensity of signal suppression. For example the appearances of the first transparent regions, due to lack of neutral hydrogen, at $\nu_{obs}\simeq\rm147\,MHz$ and angular position $\rm3.2^{\circ}$ and $\rm1.1^{\circ}$ shows that ionization around sources are more consistent for the simulation with inhomogeneity dependent clumping. This is the case even at higher frequency $\nu_{obs}>180\,MHz$ ($z<7$), during the final phases of reionization the morphology and size of the \textit{percolation cluster} strongly depends on the clumping model employed by the simulation. BHC model has  large regions of feeble emission $\sim 3$ mK that are extensively linked together. The IC model shows the same morphology but with considerably smaller and more isolated regions of signal. The SC model, in the other hand, shows a conspicuous lack of signal and regions of emission have only of a few Mpc size.

\begin{figure*}
	\includegraphics[width=\textwidth]{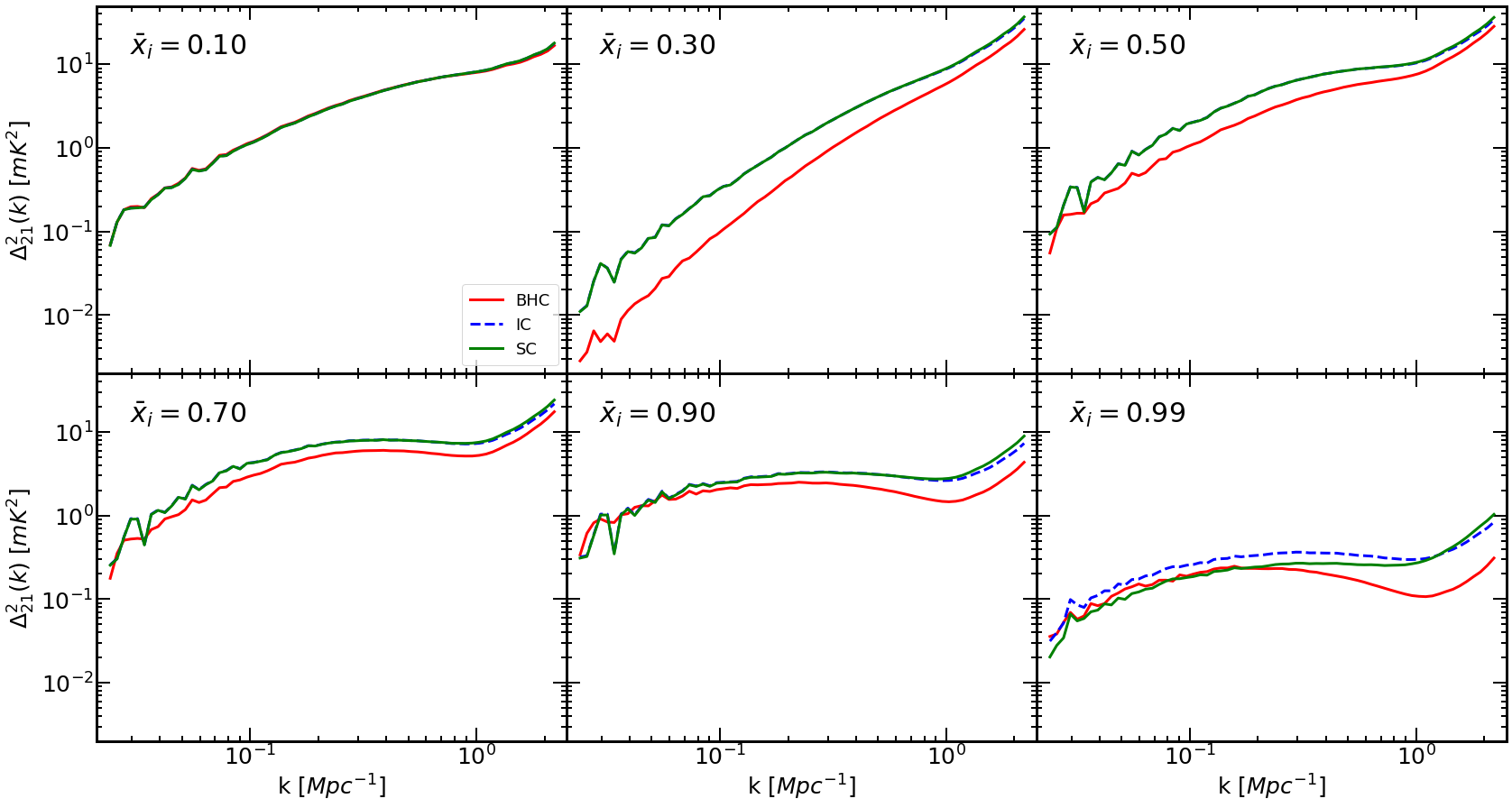}
	\caption{The effect of clumping factor on the 21-cm power spectra compared at volume ionization fraction $x_{\rm i}=0.1$, 0.3, 0.5, 0.7, 0.9 and 0.99 for the models under study: BHC (red, solid), IC (blue, dashed) and SC (green, solid).}
	\label{fig:21cmPk} 
\end{figure*}

These differences between models are more clearly observed in the statistics
of the 21-cm differential brightness temperature fluctuations.
are significant variation in the statistics of the differential brightness temperature - rms, PDFs, skewness and power spectra - shown in \autoref{fig:dTb_rms_and_mean}, \autoref{fig:21cm_PDF}, and \autoref{fig:21cmPk}.
The low frequency cut-off is chosen for range where differences between models becomes noticeable. The high density peaks get ionized early, and the corresponding H~II regions are smaller then the interferometer resolution, thus their effect on rms (\autoref{fig:dTb_rms_and_mean}, top) is to diminish the averaged $\rm\delta T_b$ without increasing fluctuations. At this stage the signal mostly follows the underlying density field, apart from the peaks and there is little difference between the models. The observed frequency of the RMS dip indicates the timing at which HII regions become larger then the interferometry smoothing scale and eventually start to overlap locally. This is the case at frequency larger then $\rm120\,MHz$ ($z\approx11$). For the IC and SC models, the turnover occurs earlier and with a steeper slope than the BHC model, indication that signal fluctuations increase faster and stronger. Moreover the peak value of the RMS fluctuations varies, in the case of IC and SC models the amplitude is $14 \%$ higher, despite having a lower averaged brightness temperature then the homogeneous case, indicating that the signal is sensitiive to a more physical treatment of the clumping factor. This is the consequence of a lower clumping factor values in under-dense regions, consistent with the conclusion in \autoref{sec:IonizHistory}. The faster propagation of I-fronts, in the vast low density regions, leads to a earlier second peak in the RMS of the two former approaches. In order of appearance at $\rm\nu_{obs}=165\,MHz$ ($\rm z=7.56$) for SC, 169 MHz ($\rm z=7.34$) for IC and slightly later at 176 MHz ($\rm z=7.06$) for BHC, respectively when the average neutral fraction is $\rm \bar{x}_n=0.33$, 0.28 and 0.25. The subsequent decline is the results of reionization reaching its final stage, with almost complete ionization.

The averaged 21-cm fluctuations level at different scales is reflected in the power spectra (\autoref{fig:21cmPk}), where we compare the results for models BHC, IC and SC at epochs at which the mean ionization fractions are $x_{\rm i}=$ 0.1, 0.3, 0.5, 0.7, 0.9, as well as around reionization completion $x_{\rm i} =$ 0.99. At first, the 21-cm signal follows the underlying density distribution of neutral hydrogen and the power spectra are very similar and approximatively a power law in all three cases. 
The flattening of the power spectra is an indication of the expanding ionized region, shifting the signal toward larger scales while suppressing small structures. Interestingly, this characteristic appears at the same scale regardless of the clumping model but modest difference in amplitude of signal. The BHC model yields systematically lower power at all scales and at all redshifts except close to overlap.
The stochastic relation between local overdensity and clumping factor does not have a large effect throughout most of reionization, and is noticeable predominantly at small scales later on. The most significant differences between IC and SC models emerges at the end of reionization ($x_{\rm i}=0.99$), where the SC model has less power on all scales, by factor of up to a few. In fact, at that time the SC model has less power than even the BHC, except at the small scales $k \rm > 0.3 Mpc^{-1}$.

The 21-cm signal fluctuations are strongly non-Gaussian \citep[e.g.][]{Mellema2006,Giri2019PdPS} and therefore are not fully described by the power spectra. We therefore also present the 21-cm differential brightness temperature distribution moments of first (PDFs; \autoref{fig:21cm_PDF}) and second order (skewness; \autoref{fig:dTb_rms_and_mean}, lower panel). For all the models and all times, 21-cm PDFs are bimodal in nature, which is a clear signature of non-Gaussianity \citep[e.g.][]{Ichikawa2010,Giri2018Optimal}. Even though all the models show non-Gaussinity, there are significant variations between models. The SC and IC models are much more non-Gaussian, with many more pixels at both high low values. Particularly, they show a very strong tail at high values. This is somewhat stronger for the SC model at all redshifts, indicating that the clumping scatter yields more high brightness temperature peaks, by factor of a few. The signal skewness confirms these observations. It is going from negative to positive symmetry at $\rm\nu_{obs}\simeq170\,MHz$, when the volume ionized fraction is close to $x_{\rm i} = 0.6-0.7$ and the RMS fluctuations reach maximum. Differences between models are noticeable only later, once the simulation overpass the peak in fluctuations, at frequency larger then 180 MHz. At this point the skewness increases exponentially.

\section{Conclusion} \label{chap:Conclusion}

Studies of the large scale reionization morphology and its imprint on the
observable signatures requires large simulated volumes of a several hundred cMpc per side. Due to computational limitations which limit the dynamic range, uniformly high resolution cannot be achieved in such a volume. Therefore no general model of the local recombinations on scale below the resolution of large numerical simulation exists. Typically a constant value of clumping factor is used, but recently we presented a more general model (\citetalias{Mao2019}), that depends on the local density, and we demonstrated how an over-simplistic treatment of the clumping factor can have a strong effect on the simulated reionization timescale, topology and size distributions of the ionized region.

In the current work we extend and improve this method by including an empirical stochastic subgrid gas clumping (SC) model (see \S\ref{SC_model}) based on the results from high-resolution N-body simulation, where the full range of relevant fluctuations is fully resolved. Our approach considers a novel parametrization of the correlation between local IGM overdensity and clumping factor, which take into accounts the scatter due to e.g. tidal forces. We employ a high resolution N-body simulation SB, of spatial resolution $\rm260\,pc$ per side, that resolve the Jeans length of the cold IGM and structure evolution on scale much smaller then the resolution of EoR simulations. The density-binned scatter is then modelled with a log-normal distribution. Those distributions are then randomly sampled to create a realization of the scatter. We then apply our method to the density fields of  larger volumes LB-1 ($714\mpc$ per side) and LB-2 ($349$~Mpc) to infer its sub-grid clumping factor (see \S\ref{sec:ClumpRealiz}). Subsequently we post-process the large scale N-body snapshot with \texttt{C$\rm^2$Ray} radiative transfer cosmic reionization simulation code, in order to present the impact of various modeling approaches for gas clumping on reionization observables (see \S\ref{chap:RTresult}). We then compare our stochastic model SC with the inhomogeneous clumping model, IC, which is a simple deterministic density-dependent fit, and a globally redshift dependent averaged clumping factor BHC, whereby the subgrid gas clumping is independent of the local density.

We find that density-dependent clumping models, IC and SC, exhibit similar behaviour for globally averaged quantities, meanwhile there is a tangible difference when compared to the volume-averaged model BHC. For instance, the reionization history (\autoref{fig:xi}) is delayed by as much as $\rm\Delta z\sim0.3$ at $\rm x_i=0.7$ ($\rm z\sim7.5$) and the average neutral fraction decrease swiftly for $\rm z<10$. The evolution of ionized regions in IC and SC models is a bit faster due to the on average lower gas clumping factor that decreases gas recombination in the under-dense regions. Meanwhile, as structure formation advance, the higher clumping factor $\rm C > 20$ in high-density regions considerably increase the recombination rate, such that recombination is twice as effective as in the BHC model case for $\rm z<12$. We find that the increase of rate in these regions, due to the different density-dependent gas clumping approach, is responsible for the divergence in the simulated observables. Despite the fact that the over-dense medium constitute a minor fraction of the box volume, compared to the vast under-dense IGM, it is responsible for the majority of recombinations. Our model and the IC method behave similarly, with only $\rm5\%$ of relative error to each other. This difference is mainly due to the broad scatter at high density in the clumping-density correlation plot (\autoref{fig:plot500}). The clumping factor for IGM in the proximity of sources, is extremely high $\rm C \sim 100$ and the introduced stochasticy can extend it to a factor of few hundreds more. Moreover, the simulated electron scattering optical depth is very similar in IC and SC models and the choice of the clumping model has little effect on the feedback of sources.

The density-dependence of the subgrid gas clumping accelerates the propagation of ionizing fronts in the low density IGM (\autoref{fig:PL_dTb}), By $\rm z < 10$ ($\rm \nu_{obs} > 130\,MHz$) the regions with low 21-cm signal around the sources are more pronounced than in the BHC case. The differences between the new stochastic approach and the IC model are minor, mostly appearing at late times ($z < 7$, $\nu_{obs} > 175$), where the SC scenario presents considerably less residual neutral gas then the other two models. These last region of neutral gas are mostly in large voids and distant from any ionizing sources, therefore our interpretation is that at lower redshift the empirical stochastic model becomes predominant in under-dense IGM, accelerating the propagation of ionizing radiation in these regions. Meanwhile, at early stages of reionization the gas recombination in high density region drives the reionization process, resulting in reduction of the ionizing photons propagating into the neutral surroundings.

We compared the simulation-derived observables at the same reionization milestones, $\rm x_i=0.1,\,0.3,\,0.5,\,0.7,\,0.9$. Compared to our previous work, the bubble-size distributions (based on both mean free path and FOF methods) do not show large variation, as an indication that the SC model does not increase the recombination rate in a way that significantly alters the morphology and sizes of the ionized regions. The same conclusion can be deduced from statistics of the 21-cm differential brightness temperature. As we demonstrated in Paper I, the density-dependent model increase the amplitude and shift the fluctuations peak position to lower frequency with a difference of approximately $\rm20\,MHz$ compared to BHC model, and just a few MHz of difference when compared to the SC model. Hence, the peak occurs at stage of reionization that differ only of few percentage $\rm\bar{x}_n \approx 0.3$ for SC and IC models and $\rm0.25$ for BHC.

The PDFs of the redshifted 21-cm distributions show some notable differences between our models. While all distributions are non-Gaussian, the IC and SC yield significantly more non-Gaussianity, with long tail of bright pixels, which is very different from the BHC model. The bright tail is longer for the SC model compared to IC, predicting many more and brighter pixels at all redshifts. 

The power spectra of the 21-cm signal (see \autoref{fig:21cmPk}) show that in early phase of reionization, the BHC scenario yields a weaker signal, when compared to density dependent models on all scales. IC and SC differ somewhat at large scale $\rm k<0.1\mpc^-1$ for $x)i=0.3-0.5$. This largely disappears
by $\rm\bar{x}_i = 0.7$. Towards the final stages of reionization ($x_i=99\%$) results for three models differ. The IC model predicts the highest signal at all scales, higher by a feactor of a few compared to SC. The BHC model signal is intermediate between them for most except the smallest scales.

The results presented here are not intended as a detailed prediction of the reionization observables, but rather a demonstration that an over-simplistic treatment of the clumping factor can have strong effect on the reionization morphology and thus on simulated observables. The widely-used BHC model, overestimates the rate at which the ionized IGM recombines, and therefore have a strong influence on the timescale of reionization, morphology of the ionized region and the intensity of the expected 21-cm signal. We demonstrated that density dependent model takes better account the cumulative effect of the clumping factor on the gas recombination rate. On the other hand, we have also shown that accounting for the scatter around the average, deterministic local density-clumping relation has only modest effects on the reionization morphology and observables, predominantly towards the end of the reionization process. This indicates that the deterministic IC model is usually sufficient except possibly around and after overlap.

The gas clumping factors presented here should be considered as an upper limit to the actual clumping since they are derived based on high-resolution N-body simulations and thus do not capture the photo-ionization feedback that would suppress small-scale density fluctuations. Consequently it overestimates the recombination rate throughout reionization. We leave a more realistic approach, that follows the feedback effects, and the complex physics of the cold gas ($\rm T < 10^4\,K$) in IGM, for future work.

\section*{Acknowledgements}
ITI was supported by the Science and Technology Facilities Council (grant numbers ST/I000976/1 and ST/T000473/1) and the Southeast Physics Network (SEPNet). MB was supported by PhD Studentship from the Science and Technology Facilities Council.
HP was supported by the World Premier International Research Center Initiative (WPI), MEXT, Japan and JSPS KAKENHI Grant Number 19K23455.
YM is supported by the National Key R\&D Program of China (Grant No. 2020SKA0110401, 2017YFB0203302, 2018YFA0404502) and the NSFC Grant (No. 11761141012, 11673014, 11821303). We acknowledge that the results of this research have been achieved using the DECI resources Kay based in Ireland at ICHEC and Cartesius based in Netherlands at SURFSara with support from the PRACE aisbl. 
KA was supported by NRF-2016R1D1A1B04935414 and NRF-2016R1A5A1013277, and appreciates APCTP and KASI for hospitality during completion of this work.
We acknowledge PRACE for awarding us access to Piz Daint facility hosted by the Swiss National Supercomputer Centre (CSCS) and the MareNostrum IV hosted by the Barcelona Supercomputing Centre (BSC). The authors gratefully acknowledge the Gauss Centre for Supercomputing e.V. (www.gauss-centre.eu) for partly funding this project by providing computing time through the John von Neumann Institute for Computing (NIC) on the GCS Supercomputer JUWELS at Juelich Supercomputing Centre (JSC).

\textit{Data Availability:} The data and codes underlying this article are available upon request, but can also be re-generated from scratch using the publicly available \texttt{CUBEP$^3$M} and \texttt{C$^2$Ray} code. The code and table of parameters for \autoref{eq:Cmean}, \ref{eq:quad} and \ref{eq:realiz_ln} presented in \S\ref{sec:ClumpMod} are available on the author's Github page: \url{https://github.com/micbia/SubgridClumping}.

\bibliographystyle{mnras}
\bibliography{body/SubgridClumping}

\appendix

\section{Comparison between old and new}\label{sec:oldnew}
In the N-body simulations used in our Paper I \citep{Mao2019}, we employed the version 1 of the \texttt{CUBEP$\rm^3$M} code, the most recent version of the code at the time. Meanwhile in this paper we employed the updated version 2 of that code, that reduces the error of the near-grid point interpolation by extending the particle-particle (PP) force calculation for a particle out to arbitrary number of cells. With the latest version, the user can therefore choose how far outside the hosting cell the PP-force is active. A detailed discussion of this update can be found in \S 7.3 of \cite{Harnois-Deraps2013}.

\begin{figure}[h]
	\hskip-7mm\includegraphics[scale=0.5]{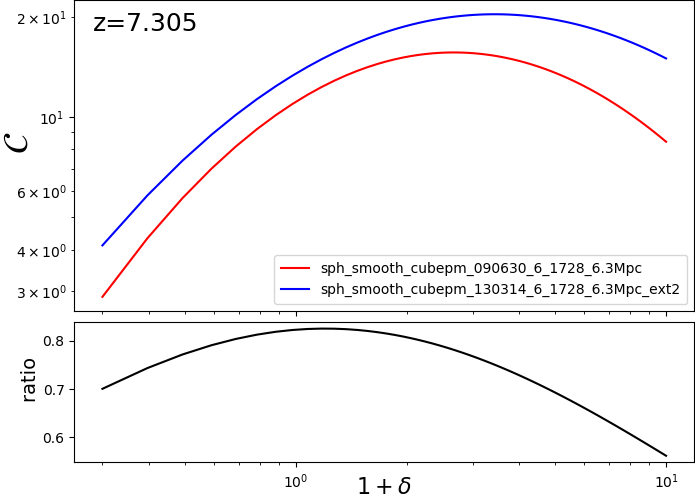} \vskip-2mm
	\caption{Correlation between local coarse IGM over-density and coarse clumping factor at redshift $\rm \rm z=7.305$ for the SB simulation. In red, the IC model interpolation ran with the version 1 of the N-body code, with the solid blue line the same quantity but with the updated code. Lower panel, the ratio between the old and new quantity.}
	\label{fig:old_new} 
\end{figure}

As an illustration of the effect of that change, in \autoref{fig:old_new}, we show the IC model of the correlation between coarse IGM over-density and coarse clumping factor at $\rm z=7.305$ for the SB simulation. In red, the interpolation obtained from N-body simulation run with first version of the code, in blue, the updated code with PP-force that extend for 2 neighbour cells. In both cases, we kept the same cosmology, initial condition and simulation parameters. In the lower panel of the figure, we show the ratio between the two old and the new result. The result of this more precise
gravity forces calculation is that the gas clumping is somewhat boosted, while
the curve retains the same shape, which has no significant effect on our method and results.

\label{lastpage}
\end{document}